\documentclass{article}
\usepackage{emulateapj,epsf}

\tighten

\lefthead{ROBINSON, GAWISER, \& SILK}
\righthead{CONSTRAINING PRIMORDIAL NON-GAUSSIANITY}

\begin{document}
\submitted{The Astrophysical Journal, 532:1-16, 2000 March 20}
\title{Constraining Primordial Non-Gaussianity with the Abundance of\\
High-Redshift Clusters}
\author{James Robinson}
\affil{Department of Astronomy, University of California, Berkeley CA 
94720;jhr@astron.berkeley.edu}
\author{Eric Gawiser\altaffilmark{1}}
\affil{Department of Physics, University of California, Berkeley CA 94720}
\altaffiltext{1}{Current address: Center for Astrophysics and Space Sciences,
 UC San Diego, La Jolla, CA 92093}
\author{and Joseph Silk}
\affil{Department of Physics, Astrophysics, 1 Keble Road, University of Oxford, OX1
3NP, UK\\and\\ Departments of Physics and Astronomy, University of
California, Berkeley, CA 94720}
\authoremail{jhr@astron.berkeley.edu}
\slugcomment{Received 1999 May 26; accepted 1999 November 11}
\begin{abstract}
We show how observations of the evolution of the galaxy cluster number
abundance can be used to constrain primordial non-Gaussianity in the
universe. We carry out a maximum likelihood analysis incorporating a
number of current datasets and accounting for a wide range of sources
of systematic error. Under the assumption of Gaussianity, the current data
prefer a universe with matter density $\Omega_m\simeq 0.3$ and are
inconsistent with $\Omega_m=1$ at the $2\sigma$ level. If we assume
$\Omega_m=1$, the predicted degree of cluster evolution is consistent
with the data for non-Gaussian models where the primordial
fluctuations have at least two times as many peaks of height $3\sigma$
or more as a Gaussian distribution does. These results are robust to almost
all sources of systematic error considered: in particular, the
$\Omega_m=1$ Gaussian case can only be reconciled with the data if a
number of systematic effects conspire to modify the analysis in the
right direction. Given an independent measurement of $\Omega_m$, the
techniques described here represent a powerful tool with which to
constrain non-Gaussianity in the primordial universe, independent of
specific details of the non-Gaussian physics. We discuss the prospects
and strategies for improving the constraints with future observations.
\end{abstract}

\section{Introduction}
Most studies of structure formation in the universe start from the
assumption that the primordial perturbations were Gaussian. Many well
motivated models, however, predict non-Gaussian fluctuations,
including topological defect models (Kibble\markcite{K76} 1976;
Vilenkin \& Shellard\markcite{VS} 1994), and certain forms of
inflation (Peebles\markcite{P83,P97,P98a,P98b} 1983,1997,1998a,1998b;
La\markcite{L91} 1991; Amendola \& Occhionero\markcite{AO91} 1993;
Amendola \& Borgani\markcite{AB93} 1994). One area which has recently
received a considerable degree of attention in the Gaussian case is
the evolution of the galaxy cluster number abundance. Several authors
(Frenk et al.\markcite{FWED} 1990; Oukbir \& Blanchard\markcite{OB92} 1992;
Oukbir \& Blanchard\markcite{OB} 1997; Fan, Bahcall \&
Cen\markcite{FBC} 1997; Gross\markcite{Getal} et al. 1998; Blanchard
\& Bartlett\markcite{BB} 1998; Eke et al.\markcite{ECFH} 1998;
Reichart et al.\markcite{Retal98} 1999; Viana
\& Liddle\markcite{VL98} 1999) have shown how cluster evolution data
can be used to constrain the matter density of the universe under the
assumption of Gaussianity, with most studies favoring a low density
universe.

Some attempts have been made to extend studies of cluster formation to
the non-Gaussian case. Oukbir, Bartlett \& Blanchard\markcite{OBB} (1997)
noted the existence of a degeneracy between the spectral index of the
primordial fluctuations and primordial non-Gaussianity. Colafrancesco,
Lucchin \& Matarrese (1989), and later Chiu, Ostriker
\& Strauss\markcite{COS} (1997) have introduced a modified version of the
Press-Schechter\markcite{PS} (1974) formalism to make predictions for
the cluster number density in a non-Gaussian model, and Robinson,
Gawiser and Silk (1998, hereafter RGS98) have generalized this
approach. Robinson \& Baker\markcite{RB} (2000, hereafter RB00) have
tested this formalism and verified that it is able to accurately fit
the evolution of the cluster number abundance observed in N-body
simulations of structure formation with non-Gaussian initial
conditions. Various authors (RGS98\markcite{RGS}; van de
Bruck\markcite{vdB} 1998; Koyama, Soda
\& Taruya 1999) have used this formalism to study the evolution
and clustering properties of galaxy clusters and thus constrain
non-Gaussianity in the primordial
fluctuations. Willick\markcite{Willick} (1999) has considered the
constraints on non-Gaussianity which can be derived from the existence
of cluster MS1054-03 at a redshift $z=0.83$. In this paper, we extend
previous work by carrying out a detailed study of cluster evolution in
the non-Gaussian case, performing a Bayesian likelihood analysis,
considering a number of different cluster observations, and accounting
for a wide range of possible sources of systematic error.

Our results demonstrate that given an independent measurement of the
matter density of the universe (which we can realistically expect to
gain in the near future from CMB and supernovae observations), cluster
evolution data can place strong constraints on non-Gaussianity. The
greatest strength of our analysis is that it uses observations to
place direct constraints on non-Gaussianity (in particular, on the
probability distribution function, or PDF, of the primordial density
field), without any reference to the details of the non-Gaussian
physics in specific models. Because of the infinite range of possible
non-Gaussian models and the difficulties in making accurate
predictions even in well specified cases, the model independent
constraints on non-Gaussianity considered here are therefore
particularly powerful. In section~\ref{sec-models} we discuss the
parameterization of our models and the process of predicting the
cluster number density at different redshifts. In
section~\ref{sec-data} we discuss the different datasets used in our
analysis and our method for computing the likelihood of observing a
dataset given one of our models. In section~\ref{sec-results} we
discuss the resulting constraints on non-Gaussianity from existing
cluster data, the likely sources of systematic error, and the
improvement we can hope to gain from future data. In
section~\ref{sec-conclusions} we draw our conclusions.

\section{Models}
\label{sec-models}
The models we consider are specified by knowledge of the power
spectrum of the fluctuations, the nature of the non-Gaussianity, the
background cosmology, and the cluster mass-temperature
relationship. We discuss the parameterization of these properties in
the following subsections.

\subsection{Power Spectrum} 
We parameterize the power spectrum $P(k)$ for each
model using a cold dark matter (CDM) form
(Bond\markcite{B91} et al. 1991), that is
\begin{equation}
P(k)\propto k T^2 (k)
\end{equation}
where 
\begin{eqnarray}
\nonumber
T(k) = \frac {\ln (1+\mbox{2.34} q)} {\mbox{2.34} q} 
\times \hspace{1.5 in}
\\
\left[
1 + \mbox{1.389}q + (\mbox{16.1} q) ^2 + ( \mbox{5.46} q) ^3 +
(\mbox{6.71} q)
 ^4 \right] ^{-1/4}
\end{eqnarray}
and 
\begin{equation}
q=\frac{k}{\Gamma h \mbox{Mpc}^{-1}}
\end{equation}
Here $\Gamma$ is the shape parameter (which is well fit by
$\Gamma=\Omega_m h \,\rm{exp}(\Omega_B-\Omega_B/\Omega_m)$ for both
open and flat CDM models, where $\Omega_m$ is the contribution of CDM
and baryons to the critical density at $z=0$, $\Omega_B$ is the
contribution of the baryons, and the Hubble constant is parameterized
via $H_0=100h$km s$^{-1}$Mpc$^{-1}$). Although this form is derived
within the context of the CDM model, it can also be used to fit power
spectra from a range of structure formation scenarios, at least over
the range of scales relevant to cluster formation. As we will see
later, our results are quite insensitive to the value of $\Gamma$,
suggesting that a more precise parameterization of the power spectrum
(for instance, allowing for the possibility of a primordial tilt) is
not necessary. We quantify the normalization of the power spectrum by
specifying $\sigma_8$, the {\it rms} density fluctuation in spheres of
radius $8h^{-1}$Mpc.

\subsection{Non-Gaussianity}
Non-Gaussianity is specified in terms of the probability distribution
function (PDF) $p_R(\delta)$ of the primordial fluctuations, where
$p_R(\delta)\,d\delta$ is the probability of the over-density
(linearly extrapolated to a redshift $z=0$) in a sphere of radius $R$
having a value between $\delta$ and $\delta+d\delta$. For most
non-Gaussian models, it is reasonable to assume that the PDF is scale
invariant over the range of scales relevant to cluster formation (see
RB00\markcite{RB} for specific examples). In this case, we can make
use of a rescaled PDF, which we denote $P(y)$, satisfying
\begin{equation}
P(y)\,dy=\frac{1}{\sigma_R} p_R(\delta) \,d\delta
\end{equation} 
where $\sigma_R$ is the {\it rms} fluctuation in spheres of radius
$R$. 

The rescaled PDF $P(y)$ has a mean of zero and an {\it rms} of
one. For a wide range of models (see RB00\markcite{RB} for examples)
$P(y)$ can be well fit by a log-normal distribution, that is
\begin{equation}
P^A_{LN}(y)=\frac{C}{\sqrt{2\pi A^2}} e^{-x^2(y)/2 - |A|x(y)}
\end{equation}
where
\begin{equation}
x(y)=\frac {\ln (B+Cy|A|/A)}{|A|}
\end{equation}
with
\begin{eqnarray}
B&=&e^{A^2/2}\\
C&=&\sqrt{B^4-B^2}
\end{eqnarray}
This distribution has one free parameter $A$, with the limit $A=0$
corresponding to Gaussianity. Following RGS98, we characterize each
PDF in terms of a single non-Gaussianity parameter $G$, where
\begin{equation}
G=2\pi \frac{\int_3^\infty P(y) dy}{\int_3^\infty e^{-y^2/2} dy}.
\end{equation}
$G$ is the probability of obtaining a peak of height 
$3\sigma$ or higher for the PDF in question, relative to that for a
Gaussian model. A value $G>1$ indicates an excess of large positive
fluctuations, $G<1$ indicates a deficit, while $G=1$ indicates
Gaussian fluctuations. Since rich clusters typically form from 
$3\sigma$ or higher peaks in the primordial fluctuations, this
parameter is a useful quantifier of non-Gaussianity in
this context.

\subsection{Background Cosmology}
The background cosmology is specified by two parameters, the fraction
of critical density at $z=0$ contributed by dark matter ($\Omega_m$) and
the fraction contributed by a cosmological constant
($\Omega_\Lambda)$. We restrict our investigation to the flat
($\Omega_m$+$\Omega_\Lambda$=1) and open ($\Omega_\Lambda=0$) cases.

\subsection{Cluster Mass-Temperature relationship}
The Press-Schechter (PS) formalism (Press
\& Schechter 1974) allows us to predict the number density of clusters
in the universe as a function of cluster mass. This formalism has been
adapted to the non-Gaussian case, by Colafrancesco,
Lucchin \& Matarrese (1989), Chiu et al.\markcite{COS} (1997), and RGS98
and has been shown in RB00\markcite{RB} to fit the cluster
evolution observed in non-Gaussian models of structure formation to
better than $25\%$ accuracy.  In particular, the number density of
clusters with masses between $M$ and $M+dM$ is given by
\begin{equation} 
n(M)dM=\frac{3f}{4\pi R(M)^3} P[y_c(M)] \frac{d[y_c(M)]}{dM}dM.
\end{equation}
Here $R(M)$ is the Lagrangian (pre-collapse) radius of a sphere giving
rise to a cluster of mass $M$, which satisfies
\begin{equation}
M=\frac{4 \pi}{3} R(M)^3 \rho_b
\end{equation}	
with $\rho_b$ being the comoving background density of the
universe. Also 
\begin{equation}
y_c(M)=\delta_c/\sigma_M
\end{equation}	
where $\sigma_M$ is the {\it rms} mass fluctuation on scale $M$,
linearly extrapolated to redshift $z$, and $\delta_c$ is the critical
overdensity for collapse, which has the value $1.69$ in a flat
universe (for fits to the weak cosmological dependence, see
Kitayama\markcite{KS} \& Suto 1996). Finally, $f$ is a correction
factor, given by
\begin{equation}
f=\frac{1}{\int_0^\infty dy P(y)}
\end{equation} 
which is included to ensure that the mass function accounts for the
entire mass of the universe. In the Gaussian case $f$ takes the value
two, as originally proposed by Press \& Schechter\markcite{PS} (1974)
and explained by Bond et al. (1991) using the excursion set formalism.

The Press-Schechter formalism allows us to accurately predict the
cluster number density as a function of cluster mass. For most of the
clusters considered in this work we observe not the mass but the
temperature of the intra-cluster medium (via the spectrum of emitted
X-rays). We therefore need to make use of a mass-temperature relation
in order to compare our models with observations. Hydrodynamical
simulations of cluster formation (Navarro, Frenk \&
White\markcite{NFW} 1995; Evrard, Metzler \& Navarro\markcite{EMN}
1996; Eke, Navarro \& Frenk\markcite{ENF} 1998) suggest that the
relationship between mass and temperature can be well fitted by
\begin{equation}
kT_{gas}=\frac{9.37}{\beta(5X+3)} \left(\frac{M} {10^{15} h^{-1}
M_\odot}\right)^{2/3} (1+z)
\left(\frac{\Omega_m}{\Omega_z}\right)^{1/3} \Delta_C^{1/3} {\rm keV}
\end{equation}
where $\Delta_C$ is the ratio of the mean halo density to the critical
density at redshift $z$ (for fitting functions in various cosmologies,
see Kitayama\markcite{KS} \& Suto 1996), $X$ is the hydrogen mass
fraction (we take $X=0.76$), and $\beta$ is the ratio of the specific
galaxy kinetic energy to the specific gas thermal energy. Numerical
simulations suggest that the average value for this quantity is in the
range $\bar{\beta}=1.0-1.3$. For the bulk of this work we take
$\bar{\beta}=1.0$, although we also investigate the effect of varying
the value. Unfortunately, knowledge of the mean value of $\beta$ is
not enough to specify our models, as the mass temperature relation is
also subject to scatter, with some clusters of a given mass being
hotter and other clusters being colder. Simulations suggest that the
spread of $\beta$ values can be reasonably fit as a log-normal
distribution, that is $\log_{10}(\beta)$ is Gaussianly distributed,
with mean $\log_{10}(\bar{\beta})$ and variance $\sigma_\beta$. Using
the distribution of $\beta$ values, we can infer the probability
$p(T|M)dT$ of a cluster of mass $M$ having a temperature between $T$ and
$T+dT$.  The expected number density $n(T)dT$ of clusters with
temperature between $T$ and $T+dT$ is then given by
\begin{equation}
n(T)dT=\int p(T|M) n(M) dM
\end{equation}
Due to the fact that $n(M)$ for typical models is a sharply falling
function of $M$, increasing the scatter $\sigma_\beta$ systematically
boosts $n(T)$. We adopt $\sigma_\beta=0.065$ as a fiducial choice,
though we also investigate the effect of variations about this
value. The additional scatter in the mass-temperature relation caused
by temperature measurement errors is typically smaller than the
intrinsic scatter discussed above, but we can account for it in the
same way by adding temperature measurement errors in quadrature to
$\sigma_\beta$. Obviously, our assumptions about the form of $P(T|M)$
will affect the inferred likelihood of observing a cluster with a
given temperature. We will see later that our results are relatively
insensitive to the amplitude of the scatter in temperature values for
a given mass, and consequently it is reasonable to assume that the
particular choice of a log-normal distribution for this scatter does
not bias our results. 

To summarize, our models are parameterized by 7 parameters: $\Gamma$,
$\sigma_8$, $G$, $\Omega_m$, $\Omega_\Lambda$, $\beta$ and
$\sigma_{\beta}$.

\section{Data}
\label{sec-data}
We extract information on the redshift evolution of clusters from
three sources:
\begin{itemize} 
\item{Henry \& Arnaud clusters:} 
This is an X-ray selected sample of clusters with galactic latitude
$b>20^\circ$ and a flux limit $f_{\rm min}=3\times10^{-11}$erg
cm$^{-2}$ s$^{-1}$ in the $2-10$ keV band (Henry \& Arnaud\markcite{HA}
1991). The sample contains 25 objects, with a median redshift
$z=0.05$, and is at least $90\%$ complete (for the purposes of this
work we shall assume $100\%$ completeness, although it makes no
difference to our conclusions).
\item{\it Markevitch clusters:}
This is a sample of clusters selected from the ROSAT All-Sky Survey
Abell cluster list (Ebeling et al. 1996) plus three known bright
non-Abell clusters, all with $b>20^\circ$ and cooling-flow corrected flux
greater than $f_{\rm min}=2\times10^{-11}$ erg cm$^{-2}$ s$^{-1}$ in
the $0.1-2.4$ keV range (Markevitch\markcite{M98} 1998). In addition,
the sample is volume limited by requiring $z_{\rm min}<z<z_{\rm max}$
with $z_{\rm min}=0.04$ and $z_{\rm max}=0.09$. For an explanation of
the cooling flow correction procedure, see Markevitch\markcite{M98}
(1998). We make use of both cooling flow corrected and uncorrected
temperatures in this analysis.
\item{\it EMSS clusters:} The EMSS clusters are selected from an X-ray survey
in the 0.3-3.5 keV band covering a total area $A_{\rm tot}=735$ square
degrees. The fraction of survey area $A$ sensitive to fluxes larger
than $f_{\rm det}$ can be fit to better than $10\%$ accuracy (Eke et
al.\markcite{ECFH} 1998) for fluxes greater than
$2.5\times10^{-13}$erg cm$^{-2}$ s$^{-1}$ by
\begin{equation} 
\frac{A(f_{\rm det})}{A_{\rm tot}}=1-3.05e^{-0.41 f_{\rm det}} +2.30
e^{-0.77f_{\rm det}}.
\end{equation}
We consider a volume limited sample of 9 clusters with $0.3<z<0.4$
whose temperatures have been measured by Henry using the ASCA
satellite (Henry\markcite{H97} 97). We also consider a sample
comprising two objects with ASCA temperatures
(MS0015.9+1609, $z=0.54$, $kT=8.0\pm0.6$ keV, Mushotzky \&
Scharf\markcite{MS} 1997; MS0451.6-0305, $z=0.54$, $kT=10.4\pm1.2$ keV,
Donahue\markcite{D96} 1996), for which we take the redshift limits to
be $0.5<z<0.65$. Finally, we consider another sample comprising one
object (MS1054.5-0321, $z=0.83$, $kT=12.3^{+3.1}_{-2.2}$ keV, Donahue et
al.\markcite{Detal} 1998), with redshift limits taken to be
$0.65<z<0.9$. We will investigate whether this choice of redshift
limits affects our results.
\end{itemize}

We analyze each sample as follows: First, we model the relationship
between luminosity ($L$) and temperature for each sample as a Gaussian
distribution in $\log(L)$, with mean given by
\begin{equation}
\log_{10}(L_X/(10^{44}\mbox{erg s}^{-1}))=A_1 \log_{10}(kT/{\rm keV}) + A_2
\end{equation}
and a standard deviation
\begin{equation}
\sigma_L=A_3  \log_{10}(kT/{\rm keV}) + A_4
\end{equation}
The subscript $X$ denotes the observing band for the sample in
question, and fiducial values for the parameters $A_1$, $A_2$, $A_3$
and $A_4$ for each of the cluster samples are (see Eke et
al.\markcite{ECFH} 1998): Markevitch $A_1=2.1$, $A_2=-1.485$,
$A_3=0.0$, $A_4=0.104$; Henry \& Arnaud $A_1=3.93$, $A_2=-2.92$,
$A_3=-0.52$, $A_4=0.70$; EMSS clusters $A_1=3.54$, $A_2=-2.85$,
$A_3=-0.47$, $A_4=0.69$.  Having assumed a luminosity-temperature
($LT$) relation, we compute the average volume $V_{\rm max}(T)$ in
which a cluster of temperature $T$ could be seen for the sample in
question:
\begin{eqnarray}
\nonumber
V_{\rm max}(T)=\int dL \,p(L|T) \int_{z_{\rm min}}^{z_{\rm max}}dz
\,\frac{d^2V} {d\omega dz}\, 
\\
\times \omega(L,z) \,\theta(L-4\pi f_{min} d_L^2 (z))  
\end{eqnarray}
where $p(L|T)$ is the probability distribution of $L$ for a given
value of $T$ (defined above), $d^2V/(d\omega dz)$ is the comoving volume
element per unit redshift per solid angle, $\omega(L,z)$ is the solid
angle associated with flux limit $L$ at redshift $z$, $d_L$ is the
luminosity distance, and $\theta(x)$ is the Heaviside function
\begin{equation}
\theta(x)=\left\{
\begin{array}{ll}
1&\ldots x\ge0\\
0&\ldots x<0
\end{array}.
\right.
\end{equation}

The $LT$ relation for each sample is subject to various
uncertainties, and at high redshifts the number of clusters for which
luminosity and temperature have been observed is so low that any
accurate determination is virtually impossible. For flux limited
samples, uncertainties in $p(L|T)$ will lead to uncertainties in
$V_{\rm max}$, and therefore to uncertainties in the inferred number
density of clusters. However, for surveys which are limited in both
flux and volume, sufficiently hot clusters will typically be so bright
that for all reasonable luminosities they could be seen throughout the
entire survey volume, so the uncertainty in $V_{\rm max}$ is much
smaller. We investigate this effect by plotting the 
cumulative temperature function
\begin{equation}
N_{>T}=\sum_{T_i>T}\frac{1}{V^i_{\rm max}}
\end{equation}
for the Markevitch and Henry samples discussed above ($N_{>T}$ is the
number density of clusters with temperature greater than $T$, $T_i$
and $V_{\rm max}^i$ are the temperature and maximum volume for the
$ith$ cluster in the sample, and the sum runs over all clusters),
using four different assumptions about the $LT$ relationship. For
``Fiducial LT'' we use the $LT$ model and parameters discussed
above. For ``High LT'' we multiply luminosities in the fiducial model
by a factor of 10, and for ``Low LT'' we divide by a factor of 2. For
``Actual Luminosities'' we revert to the method for calculating
$V^i_{\rm max}$ for the $i^{\rm th}$ cluster used by Eke et al. (1998a), that is
\begin{equation}
V^i_{\rm max}(T)=\int_{z_{\rm min}}^{z_{\rm max}}dz
\,\frac{d^2V} {d\omega dz}\, \omega(L^i,z) \,\theta(L^i-4\pi f_{min} d_L^2 (z)), 
\end{equation}
where $L^i$ is the luminosity of the $i^{\rm th}$ cluster. 

Results for these four cases in a critical density universe are shown
in Figures 1 and 2.
The
Markevitch sample (Fig. 1), which is volume
limited, shows virtually no dependence on the $LT$ relationship for
$T>6.3$ keV, and the Henry sample (Fig. 2) is
similarly robust for this temperature range. For the
remainder of this work, we will concentrate our analysis on clusters
with temperatures $T>6.3$ keV for the low redshift samples, $T>8.0$ keV
for the $0.5<z<0.65$ sample, and $T>10$ keV for the $0.65<z<0.9$
sample. As discussed, restricting our analysis to the highest
temperature clusters reduces the dependence of our results on
uncertainties in the $LT$ relation. As a consequence, we will not
worry about the effect of $K$-corrections, whose typical effect on the
luminosity is less that the uncertainties in the $LT$ relation. An
additional advantage of restricting our study to the highest
temperature clusters is that heating of the intra-cluster medium by
supernovae (which could affect the cluster mass-temperature relation)
is likely to be least important in this regime, as discussed by
Viana\markcite{VL98} \& Liddle (1999).

\vbox{%
\begin{center}
\leavevmode
\hbox{%
\epsfxsize=7.5cm
\epsffile{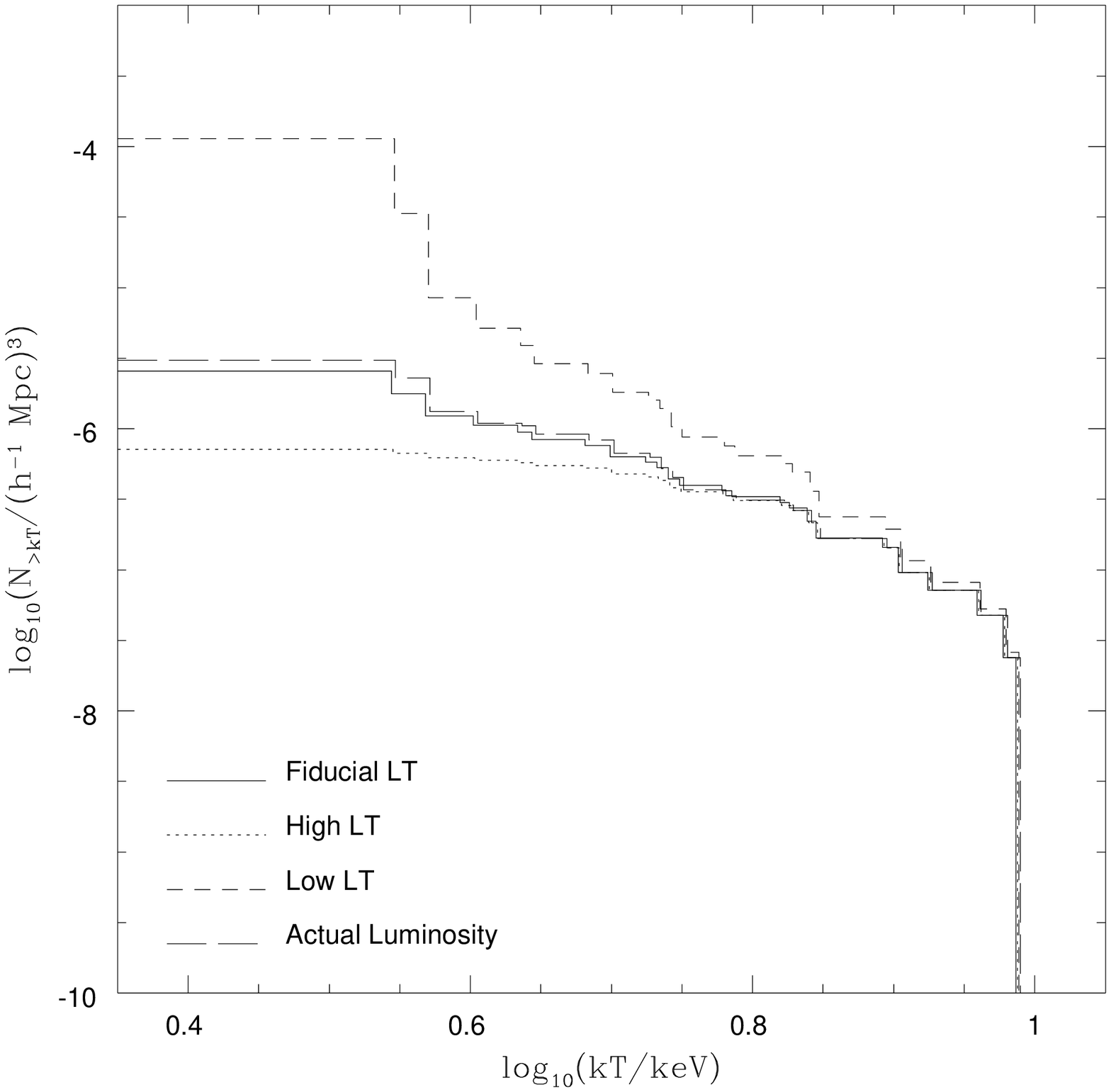}}
\begin{small}
\figcaption{\small 
Cumulative temperature function for the Markevitch (1998) cluster
sample, assuming $\Omega_m=1$, $\Omega_\Lambda=0$, and four different
versions of the luminosity-temperature relation, as discussed in the
text. }
\label{fig-markevitch_lt}
\end{small}
\end{center}}

We comment briefly on the adopted slope of the $LT$ relation. For the
Markevitch sample, we adopt a slope $A_1=2.1$ as measured by
Markevitch (1998) for the 0.1-2.4 keV observing band. This corresponds
to a slope for the bolometric luminosity vs temperature relation of
$2.64$. We follow Eke et al.\markcite{ECFH} (1998) in adopting a slope
$A_1=3.93$ for the Henry and Arnaud sample, and $A_1=3.54$ for the
EMSS clusters. Self-similar scaling arguments on the other hand would
imply a slope $A_1=2$ (Kaiser\markcite{K86} 1986), while a different
study of the low redshift LT relation (Arnaud \& Evrard\markcite{AE}
1999) finds a slope of $2.88$. As discussed, our analysis is not
strongly affected by uncertainties in the $LT$ relation, and
variations of the slope within the range suggested by the observations
do not significantly alter our conclusions.

\vbox{%
\begin{center}
\leavevmode
\hbox{%
\epsfxsize=7.5cm
\epsffile{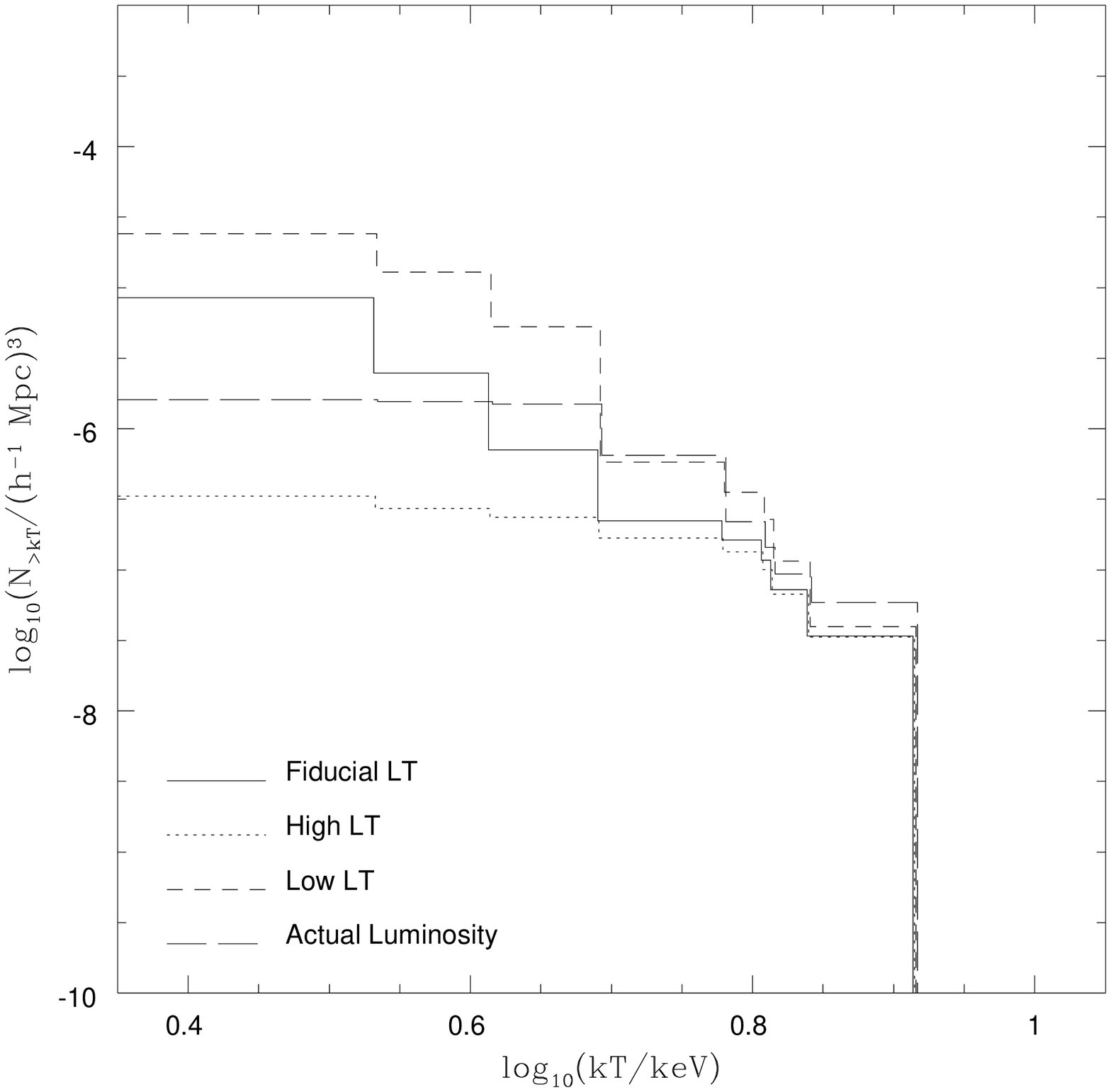}}
\begin{small}
\figcaption{\small 
Cumulative temperature function for the Henry (1997) cluster
sample, assuming $\Omega_m=1$, $\Omega_\Lambda=0$, and four different
versions of the luminosity-temperature relation, as discussed in the
text.  }
\end{small}
\label{fig-henry_lt}
\end{center}}

From the above discussion we anticipate that for some range of
temperatures $T_{\rm min}< T < T_{\rm max}$, the maximum volume
$V_{\rm max}$ will be robust, depending only weakly on uncertainties
in the $LT$ relationship. Exploiting this fact, we can make use of a
simple expression for the likelihood function: assuming that the
clusters are Poisson distributed, the likelihood ${\cal L}$ of
observing a sample of $N_{\rm c}$ clusters given a temperature
function $n(T)$ is
\begin{equation}
{\cal L}=\prod_{i=1\ldots N_{\rm c}} n(T_i) V_{\rm max} (T_i) 
\prod_{j=1\ldots N_{\rm b}-N_{\rm c}} (1-n(T_j) V_{\rm max} (T_j))  
\end{equation}
where the temperature range has been divided into a set of $N_{\rm b}$
bins, the sum $i$ runs over all bins containing a cluster, the sum $j$
runs over all bins not containing a cluster, and the bins are
sufficiently small that the probability of finding more than one
cluster in a bin is negligible. In the limit that the bin size tends
to zero, $\ln{\cal L}$ satisfies (up to an additive constant ${\cal
L}_0$)
\begin{equation}
\ln{\cal L} = \sum_i \ln\left(n(T_i) V_{\rm max} (T_i)\right) -
\int_{T_{\rm min}} ^{T_{\rm max}} n(T) V_{\rm max} (T) dT + {\cal L}_0
\end{equation}
where the index $i$ runs over all clusters in the sample satisfying
$T_{\rm min}<T_i<T_{\rm max}$.  Provided $V_{\rm max}$ is well
determined for the range of temperatures we will consider, this
expression gives a robust determination of the likelihood. Eke et
al.\markcite{ECFH} (1998) have used a very similar expression for the
likelihood which also takes account of individual redshifts within a
cluster sample. However, since the $n(T)$ laws do not change too
dramatically within the redshift ranges probed by each of our samples, we
choose instead to work just with $n(T)$ for the median redshift of
each sample.

\section{Results}
\label{sec-results}
\subsection{Current Data}
We now consider in detail the constraints we can place on $\Omega_m$
and non-Gaussianity using current observational data. As we will see
later, the results do not depend strongly on whether the universe is
flat or open, so for the time being we will concentrate on the case of
an open universe with zero cosmological constant. To start our
discussion, we consider the constraints placed by just two of the
datasets: the Markevitch data at low redshift, and the $0.65<z<0.9$
data containing MS1054.5-0321 at high redshift. Figure 3
shows confidence limits derived by combining these two datasets: in
the $\Omega_m-\sigma_8$ plane assuming an open universe with $G=1$
(left panel), in the $G-\sigma_8$ plane assuming $\Omega_m=1$ (central
panel) and in the $G-\sigma_8$ plane assuming $\Omega_m=0.3$ (right
panel). The solid lines show 90$\%$ confidence limits for the
Markevitch data, the dotted lines show 90$\%$ limits for the
$0.65<z<0.9$ data, and the dark and light shaded regions show 68$\%$
and 90$\%$ confidence regions for the combined data sets. Confidence
limits are computed assuming uniform priors in all variables. Under
the assumption of Gaussianity, the combined datasets prefer
$\Omega_m\simeq 0.3$ with $\Omega_m=1$ excluded at more than the
$2\sigma$ level, though we will see in a moment that various
systematic uncertainties might affect this conclusion. A critical
universe is consistent with the two datasets if the fluctuations
are non-Gaussian with $G\gtrsim 2.5$.

\vbox{%
\begin{center}
\leavevmode
\hbox{%
\epsfxsize=7.5cm
\epsffile{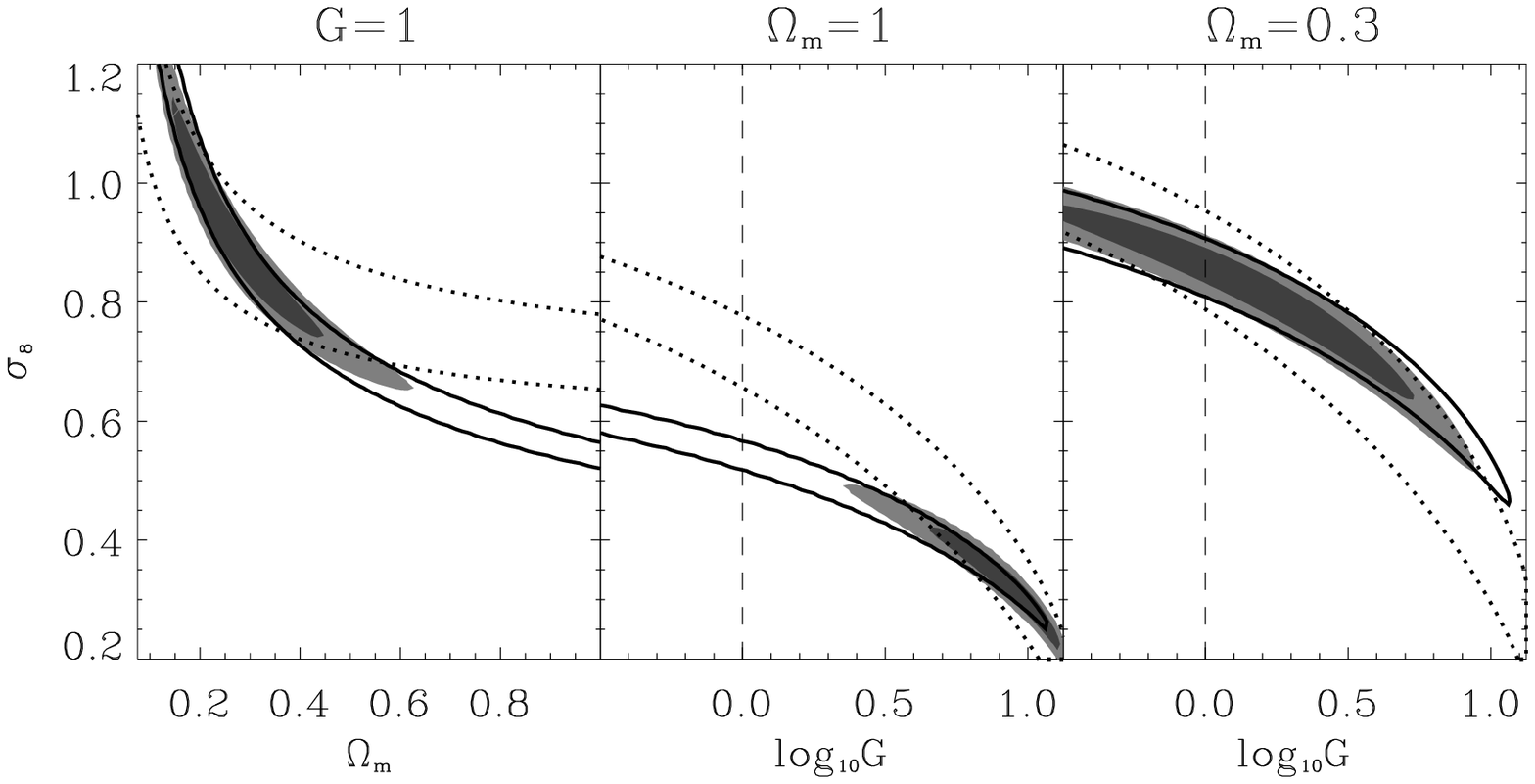}}
\begin{small}
\figcaption{\small 
Confidence limits from the Markevitch data (solid lines, 90$\%$
limits), the $0.65<z<0.9$ data (dotted lines, 90$\%$ limits), and for
the combined data sets (dark shaded region, 68$\%$ limits; light
shaded region, 95$\%$ limits), for the cases $G=1$ (left panel),
$\Omega_m=1$ (central panel) and $\Omega_m=0.3$ (right panel). The
vertical long-dashed lines in the center and right panels show the
location of Gaussian fluctuations.}
\end{small}
\label{fig-square}
\end{center}}

The discussion above, which uses just two of our datasets, illustrates
the basic degeneracy (in the $\Omega_m-\sigma_8$ or $G-\sigma_8$
plane) intrinsic to cluster observations at a single redshift, and how
this degeneracy is broken using observations of cluster
evolution. Before embarking on a full likelihood analysis
incorporating all of the datasets, we will check the consistency of
the various datasets and investigate various systematic effects which
might alter our conclusions. 

\vbox{%
\begin{center}
\leavevmode
\hbox{%
\epsfxsize=7.5cm
\epsffile{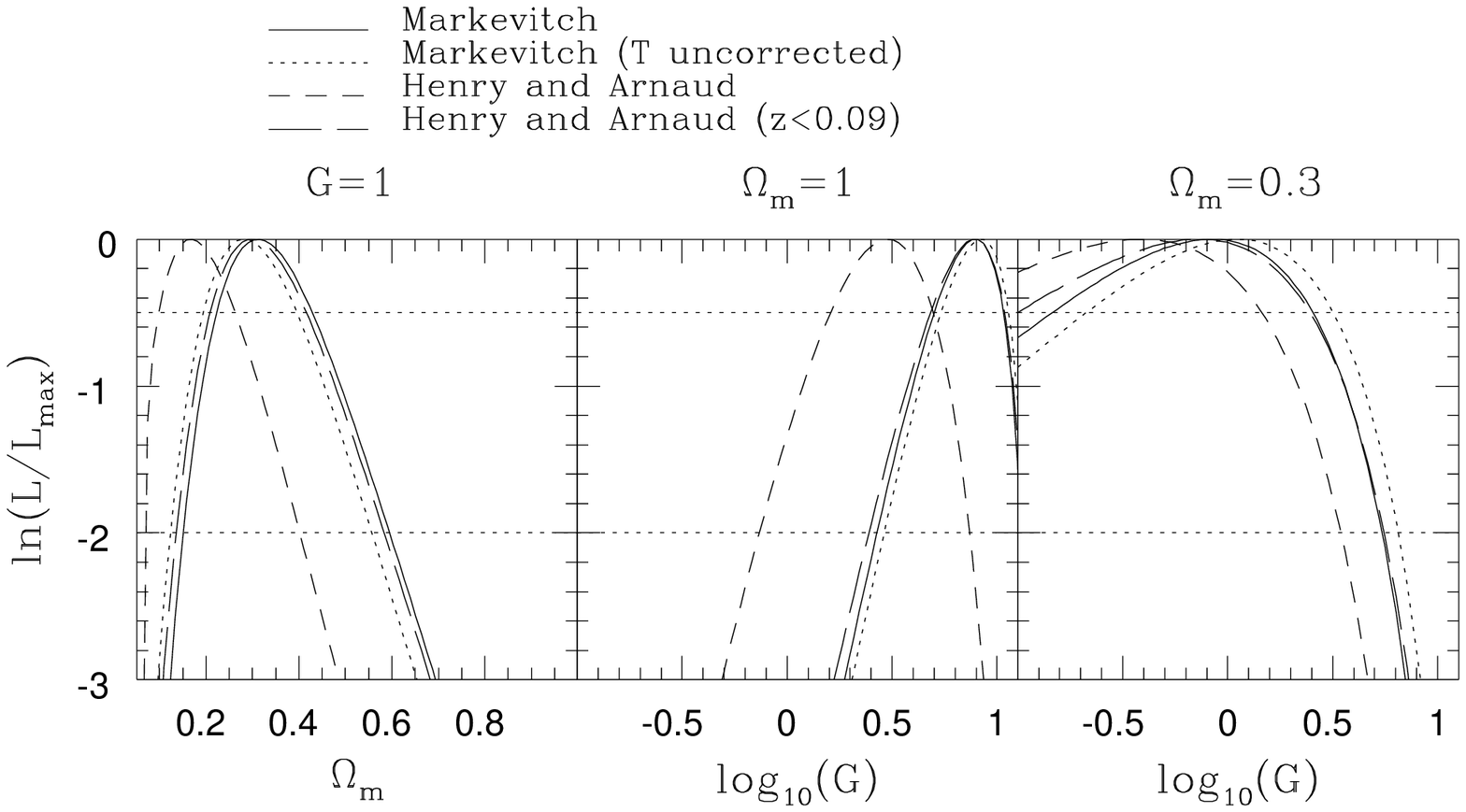}}
\begin{small}
\figcaption{\small 
Likelihood (marginalized over $\sigma_8$) for the $0.65<z<0.9$ data
combined with different low redshift data, as labeled. The dotted
horizontal lines in this and the following figures 
show likelihood thresholds corresponding to $1\sigma$
(top line) and $2\sigma$ (bottom line)  confidence limits, 
in the approximation that the likelihood function is Gaussian.}
\end{small}
\label{fig-all_low_z}
\end{center}}

First, we consider the effect of using different low redshift
data. Figure 4
shows the likelihood (marginalized over
$\sigma_8$ assuming a uniform prior) for four different choices of low
redshift data, for the cases $G=1$, $\Omega_m=1$, and
$\Omega_m=0.3$. The low redshift data considered are the Markevitch
sample, the Markevitch sample with temperatures not corrected for
cooling flows, the Henry and Arnaud sample, and the Henry and Arnaud
sample with a redshift limit of $z<0.09$ imposed.  
The likelihood
functions are almost identical (in particular, the temperature
corrections in the Markevitch sample make very little difference),
with the exception of the Henry and Arnaud data without a volume
limit. The difference in this case arises because the hottest clusters
in the sample are luminous enough to be visible at redshifts larger
than $z=0.09$. For instance, the hottest cluster in the sample (A2142)
is sufficiently luminous to be detected out to a redshift of 0.014,
implying $V_{\rm max}=1.6\times 10^8 h^{-3}$Mpc$^{3}$, while the
imposition of the redshift limit reduces $V_{\rm max}$ to $4.6\times
10^7 h^{-3}$Mpc$^{3}$. Similar changes in $V_{\rm max}$ apply to the
two next most luminous clusters, implying that the three brightest
clusters are all found within the closest $1/3$ of the volume in which
they could be detected. The fact that no clusters are actually
observed at $z>0.09$ in this sample is either a chance occurrence (the
event has moderate statistical significance) or suggests that either
the luminosity-temperature relation or the number abundance of
clusters evolves significantly beyond $z=0.09$. The latter two effects
would invalidate our analysis, since we do not take into account any
evolution within a single redshift bin. However, this problem can be
eliminated by making sure the redshift bin is sufficiently small, for
instance by limiting the maximum redshift. The imposition of a volume
limit does indeed yield results which are in extremely close agreement
with those from the Markevitch sample, and we make use of the
Markevitch data to normalize the low redshift cluster abundance for
most of the remaining discussion. Later on we will check whether
using the Henry \& Arnaud or the Markevitch sample makes a significant
difference to our final results.

\vbox{%
\begin{center}
\leavevmode
\hbox{%
\epsfxsize=7.5cm
\epsffile{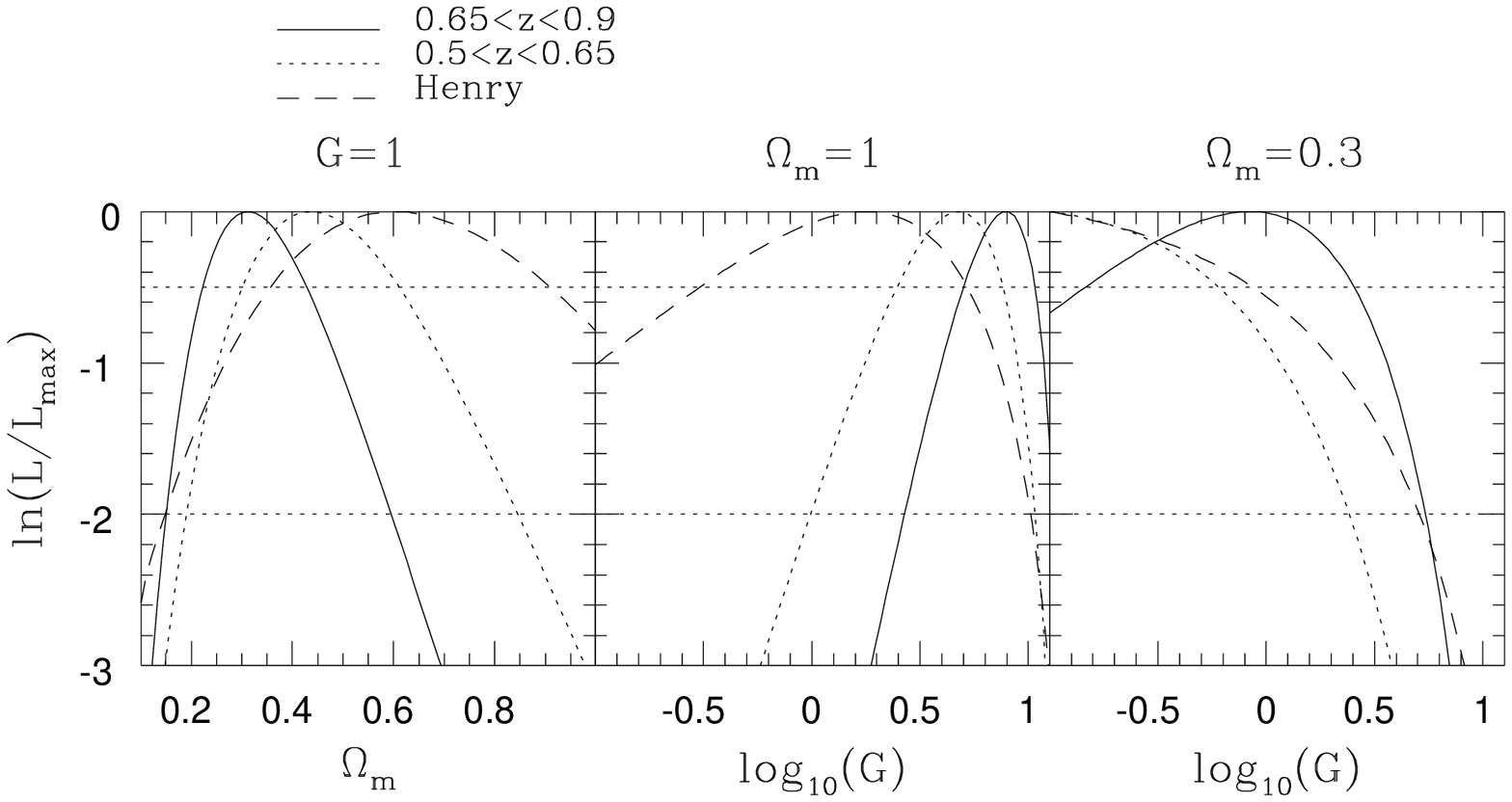}}
\begin{small}
\figcaption{\small 
Likelihood (marginalized over $\sigma_8$) for the Markevitch data,
combined with different high redshift data, as labeled. 
}
\end{small}
\label{fig-all_high_z}
\end{center}}

Second, we consider the effect of using different high redshift
datasets. Figure 5
shows the likelihood (marginalized
over $\sigma_8$) for three different choices of high redshift data,
for the cases $G=1$, $\Omega_m=1$, and $\Omega_m=0.3$. The Henry data
contains the most clusters and is the best understood systematically,
as it is complete to a known flux limit. However, even the most
extreme models are not expected to show very much evolution at this
moderate redshift, and very little of the available parameter space is
ruled out at the $2\sigma$ level. The constraints from the two
clusters in the interval $0.5<z<0.65$ are somewhat stronger, with
$\Omega_m=1$ with Gaussian fluctuations being inconsistent at the
$2\sigma$ level. However, the greatest statistical weight comes from
the highest redshift sample, despite the fact that it contains only
one cluster. (For most of the remaining discussion we consider a model
to be inconsistent at the $2\sigma$ level if $\ln(L/L_{\rm max})<-2$,
and at the $1\sigma$ level if $\ln(L/L_{\rm max})<-0.5$, which are the
thresholds which would apply if the likelihood function were Gaussian
-- most of our likelihood functions are close to Gaussian at their
peak, so this approximation should be reasonable. These thresholds are
shown by horizontal lines in our likelihood plots).

Next we investigate the effect of uncertainties in the $LT$ relation
on the allowed ranges of $\Omega_m$ and $G$.  Figure 6
shows the likelihood for a variety of choices of $LT$ relationship at
low and high redshifts. As noted, we have attempted to restrict our
analysis to temperatures which are sufficiently high that volume
limits are more important than flux limits, thus minimizing the
importance of the $LT$ relation. We see from the figures that changing
the $LT$ relation at high redshift has relatively little effect on the
likelihood function. The only change that makes a significant
difference is reducing the amplitude of the LT relationship at low
redshift. Such a change would imply that the true number density of
clusters of a given temperature is in fact higher, increasing the
inferred level of cluster evolution and increasing the preferred value
of $\Omega_m$ (or reducing the preferred value of $G$ in the
non-Gaussian case). However, even for the extreme case considered
here, where the mean luminosity is reduced by a factor of 2, the
changes to the likelihood function are not large.

\vbox{%
\begin{center}
\leavevmode
\hbox{%
\epsfxsize=7.5cm
\epsffile{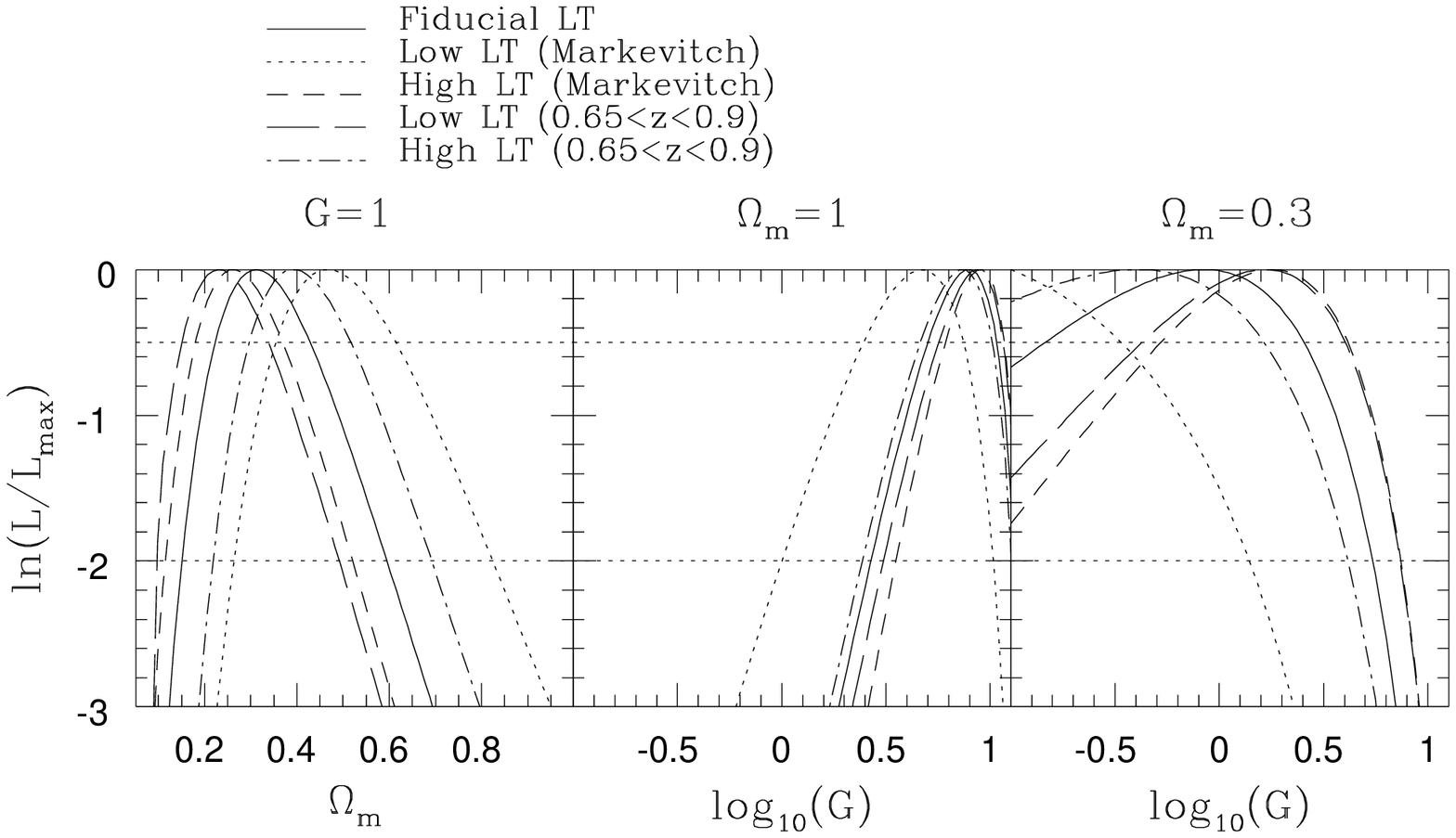}}
\begin{small}
\figcaption{\small 
Likelihood (marginalized over $\sigma_8$) for the Markevitch data and
the $0.65<z<0.9$ data under different assumptions about the cluster
luminosity-temperature relationship, as discussed in the text. 
}
\end{small}
\label{fig-all_lt}
\end{center}}

\vbox{%
\begin{center}
\leavevmode
\hbox{%
\epsfxsize=7.5cm
\epsffile{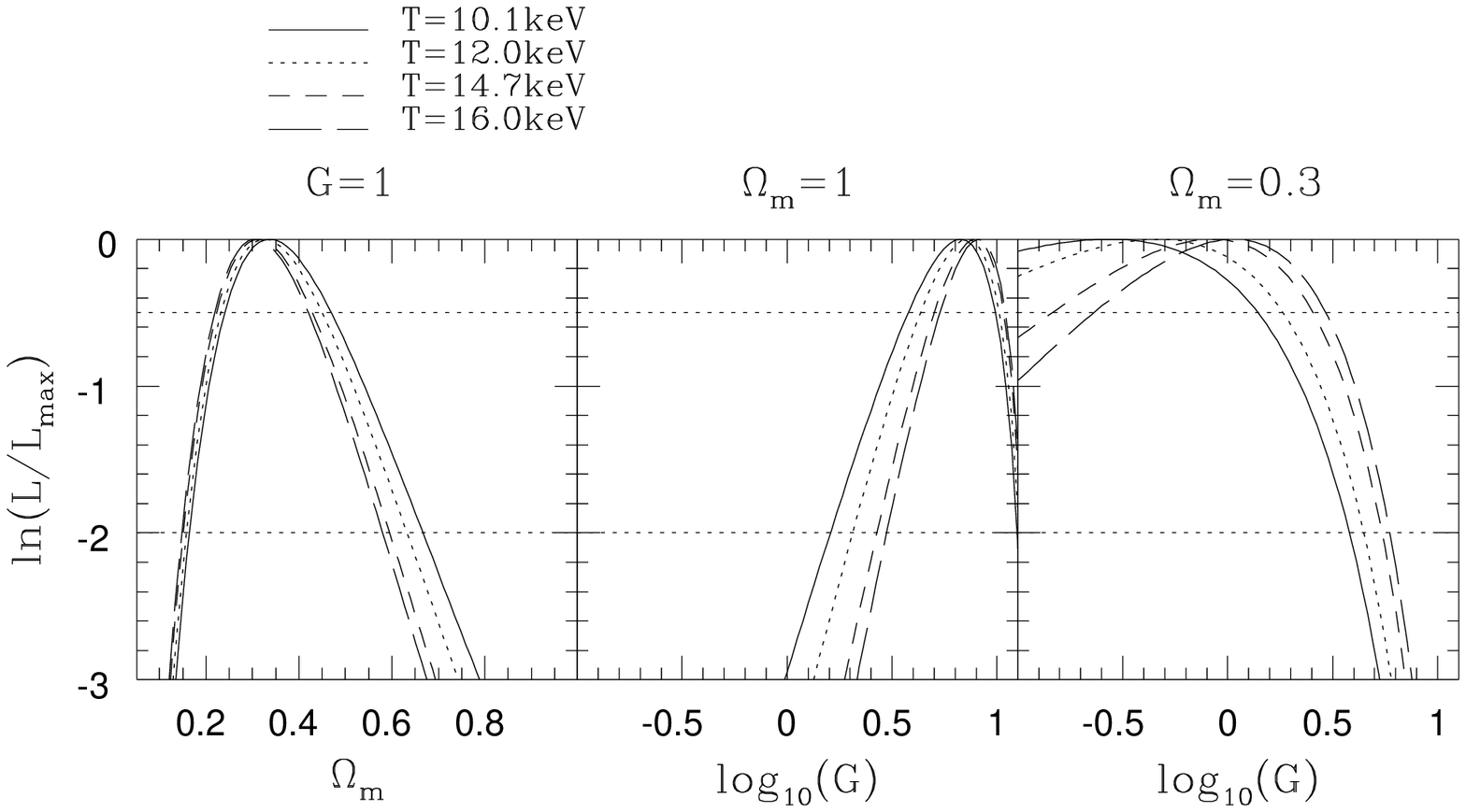}}
\begin{small}
\figcaption{\small 
Likelihood (marginalized over $\sigma_8$) for the Markevitch data and
the $0.65<z<0.9$ data, under different assumptions about the
temperature of MS1054.5-0321. The dotted
horizontal lines show likelihood thresholds corresponding to $1\sigma$
(top line) and $2\sigma$ (bottom line)  confidence limits.
}
\end{small}
\label{fig-all_lowT}
\end{center}}

Now we investigate the effect of uncertainties in the measured
temperature of the high redshift cluster MS1054.5-0321. In
Figure 7
we show the likelihood function marginalized
over $\sigma_8$ for the cases $G=1$, $\Omega_m=1$, and $\Omega_m=0.3$
under four different assumptions about the true temperature of the
cluster ($T=10.0$ keV, $T=12.0$ keV, $T=14.7$ keV and
$T=16.0$ keV). Assuming an $M_{<R}\propto R^{0.64}$ cluster profile,
the corresponding masses within a 1.5Mpc comoving radius are
$0.62\times 10^{15} h^{-1} M_\odot$, $0.76\times 10^{15} h^{-1}
M_\odot$, $0.96\times 10^{15} h^{-1} M_\odot$ and $1.06\times 10^{15}
h^{-1} M_\odot$ respectively if $\Omega_m=1$, and $0.70\times 10^{15}
h^{-1} M_\odot$, $0.86\times 10^{15} h^{-1} M_\odot$, $1.09\times
10^{15} h^{-1} M_\odot$ and $1.20\times 10^{15} h^{-1} M_\odot$
respectively if $\Omega_m=0.3$. These possibilities span the range
allowed by other estimates of the mass of the cluster from weak
lensing and velocity dispersion observations (Bahcall \&
Fan\markcite{BF} 1998a). For lower cluster temperatures, the preferred
value of $\Omega_m$ is higher (or in the non-Gaussian case the
preferred value of $G$ is higher). However, even for the lowest value
of $T$ considered, an $\Omega_m=1$ universe with Gaussian fluctuations
is inconsistent at the $2\sigma$ level, with the best fit value of
$\Omega$ (or $G$ in the non-Gaussian case) almost unchanged.

Next we consider the effect of varying the range of temperatures
$T_{\rm min}<T<T_{\rm max}$ over which we perform our likelihood
analysis for the high redshift sample. In Figure 8
we show the likelihood function marginalized over $\sigma_8$ for the
cases $G=1$, $\Omega_m=1$, and $\Omega_m=0.3$ for $T_{\rm
min}=6.3$ keV, $T_{\rm min}=10.0$ keV, and $T_{\rm min}=12.0$ keV.  For
lower values of $T_{\rm min}$ it becomes increasingly more likely that
the $0.65<z<0.9$ sample is incomplete, as there are fainter clusters
in this redshift range whose temperatures have not yet been
measured. Our analysis assuming $T_{\rm min}=6.3$ keV probably
underestimates the true number of clusters in this temperature range
(which it takes to be one), and therefore over-estimates the degree of
cluster evolution, consequently over-estimating $\Omega_m$ (or
under-estimating $G$ in the Gaussian case). Our analysis assuming
$T_{\rm min}=12$ keV is less likely to be subject to
incompleteness. However, in the case that the true temperature of
MS1054.5-0321 turned out to be less than 12 keV, our sample would have
over-estimated the number of clusters within the temperature range (as
one instead of zero), making the analysis invalid. The case $T_{\rm
min}=10.0$ keV lies in the middle, and the two cases just discussed can
be considered to represent a wide range of systematic error from
incompleteness of the high redshift sample. In the Gaussian case, the
best fit value of $\Omega_m$ is in the range $\Omega_m=0.3\pm 0.15$,
but no choice of $T_{\rm min}$ is able to reconcile a Gaussian
$\Omega_m=1$ universe with the data at the $2\sigma$ level.

\vbox{%
\begin{center}
\leavevmode
\hbox{%
\epsfxsize=7.5cm
\epsffile{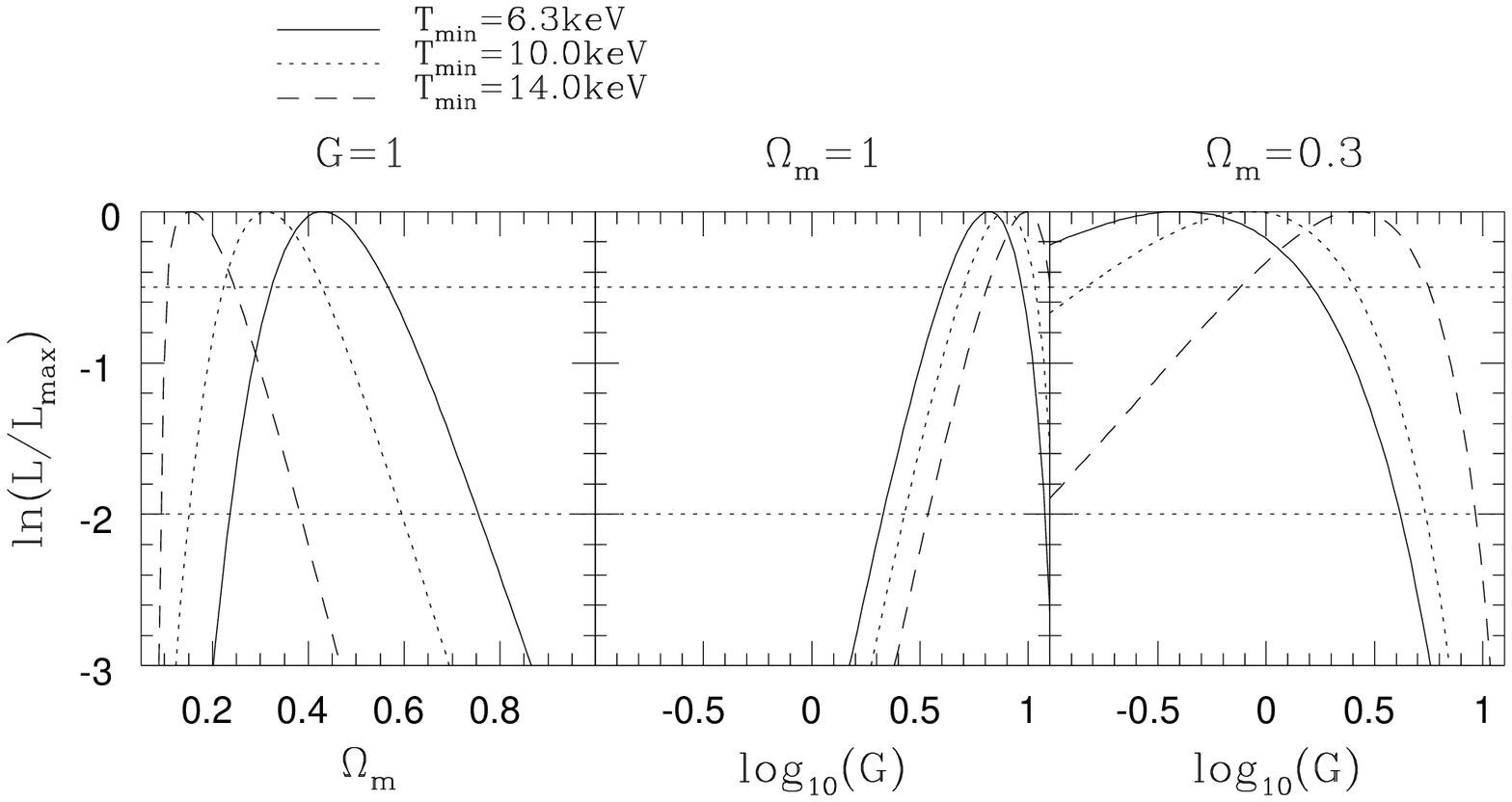}}
\begin{small}
\figcaption{\small 
Likelihood (marginalized over $\sigma_8$) for the Markevitch data and
the $0.65<z<0.9$ data, for different values of the parameter $T_{\rm
min}$. The dotted
horizontal lines show likelihood thresholds corresponding to $1\sigma$
(top line) and $2\sigma$ (bottom line)  confidence limits.}
\end{small}
\label{fig-all_tmin}
\end{center}}

Next we consider the effect of changes in the mass temperature
relationship on our likelihood analysis. Figure 9
shows the likelihood function marginalized over $\sigma_8$ for the cases
$G=1$, $\Omega_m=1$, and $\Omega_m=0.3$ for different values of the
parameters $\bar{\beta}$ and $\sigma_\beta$.  The maximum likelihood
values and exclusion limits are not significantly changed by any of
these modifications. Also, we consider the effect of varying the power
spectrum of the fluctuations. Figure 10 
shows the
likelihood function marginalized over $\sigma_8$ for the cases $G=1$,
$\Omega_m=1$, and $\Omega_m=0.3$ and a CDM shape parameter $\Gamma$ in the
range $0.05<\Gamma<0.5$, wide enough to encompass any viable
model. Exclusion limits are not significantly altered by variations of
$\Gamma$ in this range.

\vbox{%
\begin{center}
\leavevmode
\hbox{%
\epsfxsize=7.5cm
\epsffile{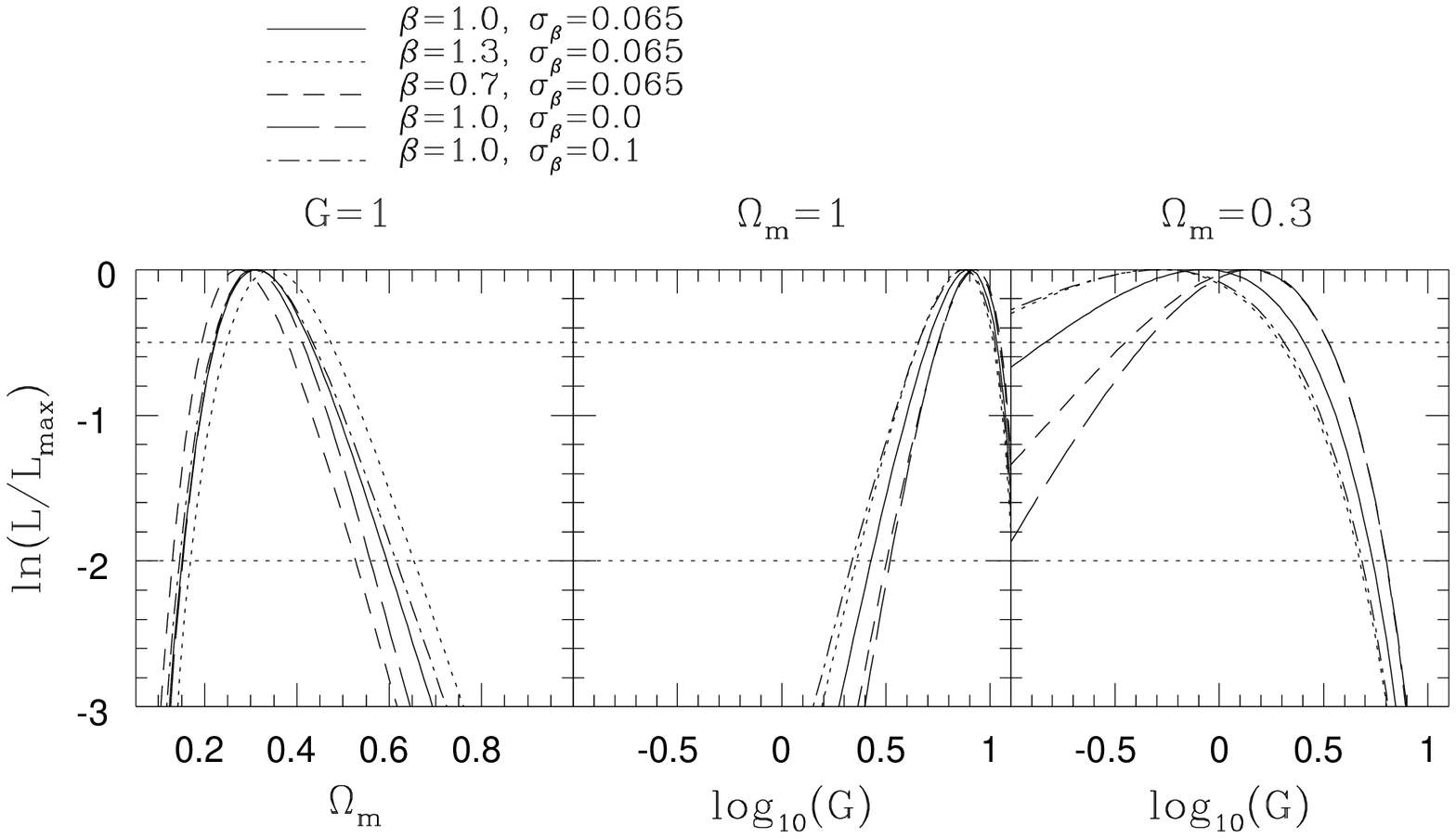}}
\begin{small}
\figcaption{\small 
Likelihood (marginalized over $\sigma_8$) for the Markevitch data and
the $0.65<z<0.9$ data, for different versions of the mass-temperature
relationship.}
\end{small}
\label{fig-all_beta}
\end{center}}

\vbox{%
\begin{center}
\leavevmode
\hbox{%
\epsfxsize=7.5cm
\epsffile{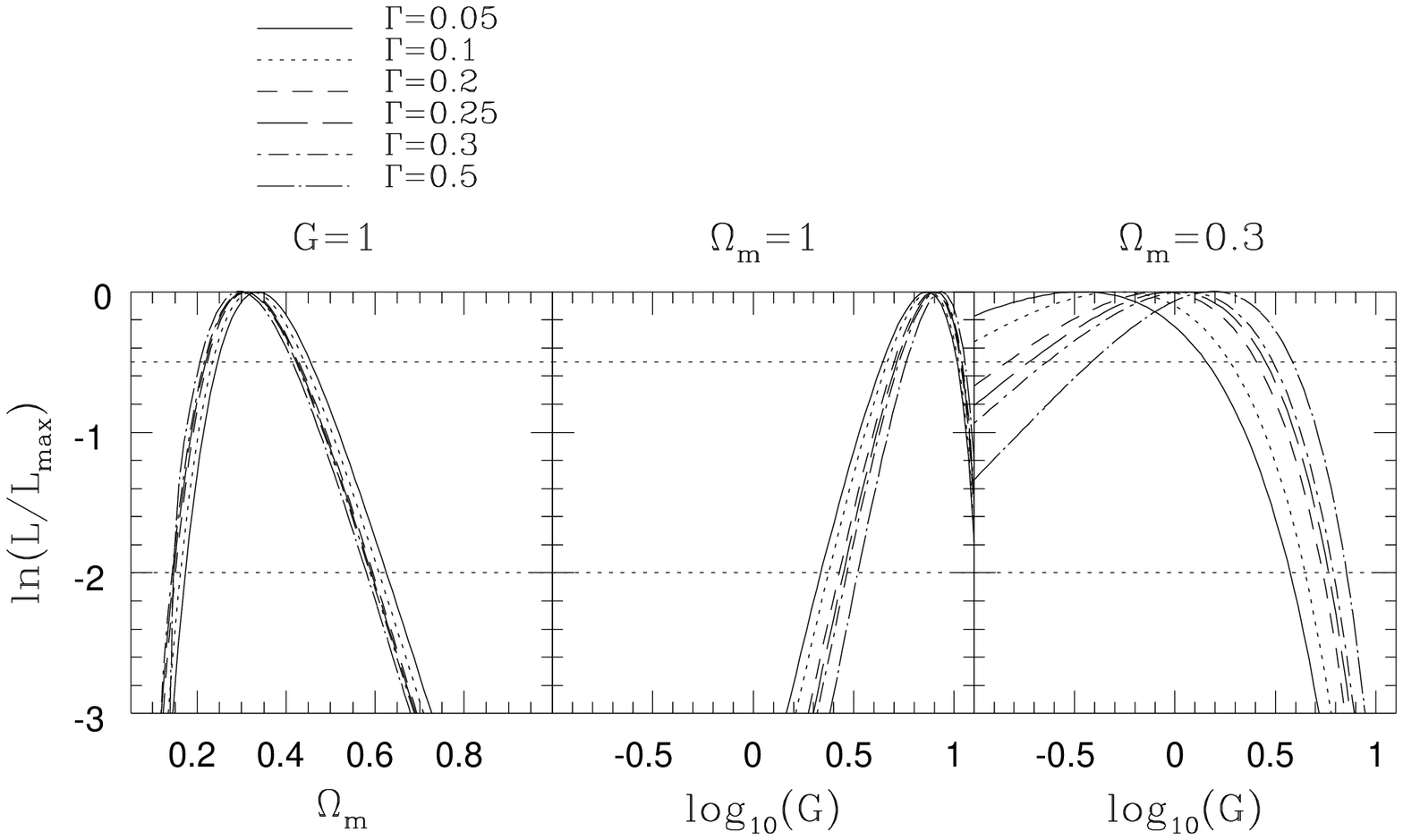}}
\begin{small}
\figcaption{\small 
Likelihood (marginalized over $\sigma_8$) for the Markevitch data and
the $0.65<z<0.9$ data, for different values of the power spectrum
shape parameter $\Gamma$.}
\end{small}
\label{fig-all_gamma}
\end{center}}

\subsection{Tests of Methodology}

Now we consider the effect of possible systematic errors in the
modified Press-Schechter prediction for non-Gaussian models. In
RB00\markcite{RB} it has been shown that the Press-Schechter formula
can model the number abundance of clusters observed in N-body
simulations with non-Gaussian initial conditions to better than 25$\%$
accuracy. We test the effect of this degree of systematic uncertainty
by multiplying the low and high redshift predictions of the cluster
number abundance by factors $F_1$ and $F_2$
respectively. Figure 11
shows the likelihood function
marginalized over $\sigma_8$ for the standard choice $F_1=F_2=1$, a
model with enhanced evolution for which $F_1=1.25$ and $F_2=0.75$, and
a model with less evolution, for which $F_1=0.75$ and
$F_2=1.25$. Increasing the degree of evolution slightly increases the
preferred value of $\Omega_m$, or reduces the preferred value of $G$
in the non Gaussian case, and decreasing the degree of evolution has a
small effect in the opposite direction. Even in this extreme case
however, where we have changed the relative number densities at low
and high redshift by a factor of more than 50$\%$, considerably more
than the uncertainty suggested by the simulations of RB00, there is little effect on the results of our likelihood
analysis.

\vbox{%
\begin{center}
\leavevmode
\hbox{%
\epsfxsize=7.5cm
\epsffile{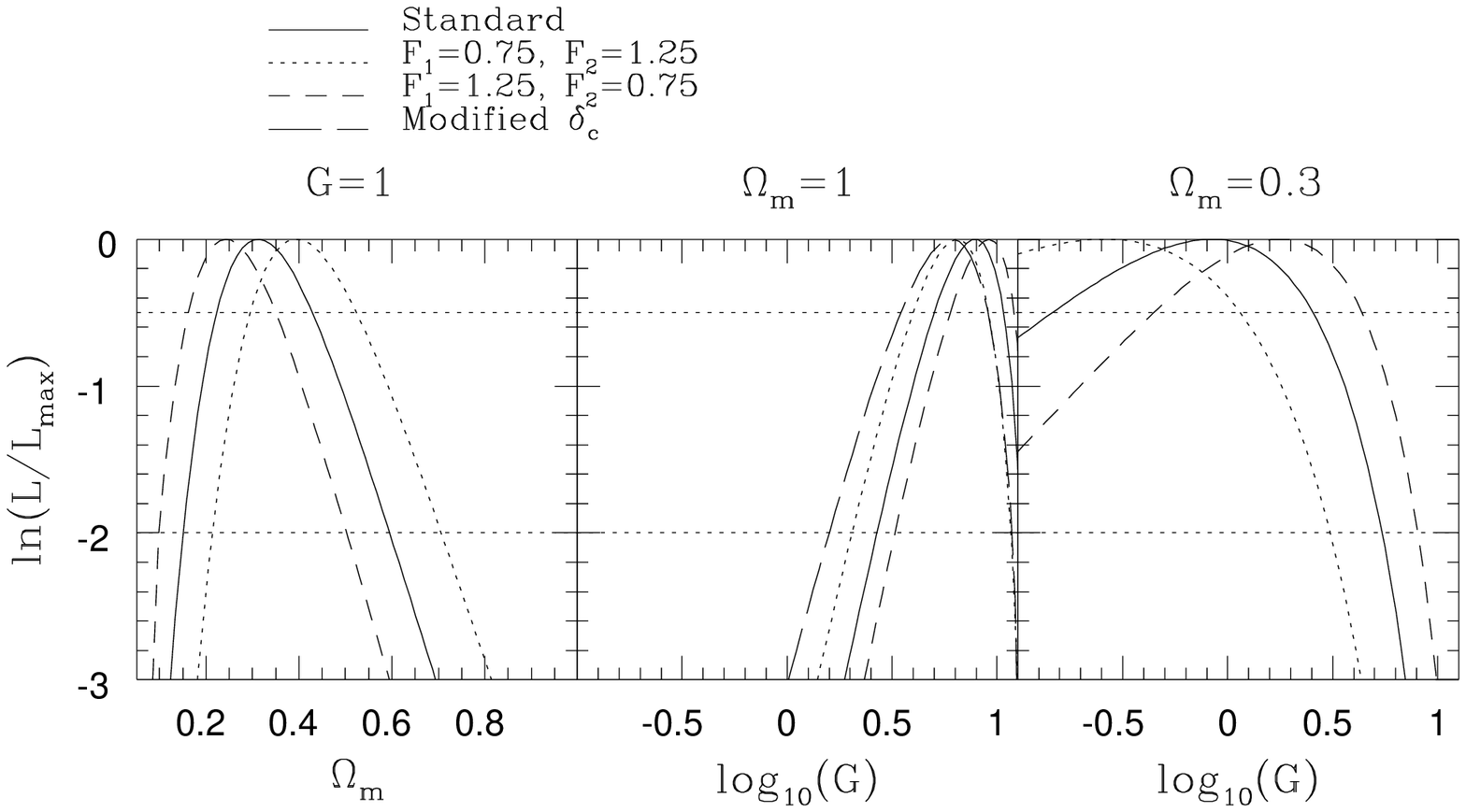}}
\begin{small}
\figcaption{\small 
Likelihood (marginalized over $\sigma_8$) for the Markevitch data and
the $0.65<z<0.9$ data, for different values of the parameters $F_1$
and $F_2$, included to account for possible systematic errors in the
Press-Schechter prediction for non-Gaussian models.}
\end{small}
\label{fig-all_error}
\end{center}}

One additional source of systematic error is the possibility that the
Press-Schechter formalism could break down in the regime of very rare
peaks. In RB00\markcite{RB} the non-Gaussian Press-Schechter formalism
has been tested in $100h^{-1}$Mpc simulations, where limited volume
restricts the possibility of testing the density of the rarest
clusters. For the case of Gaussian fluctuations, Governato et
al.\markcite{Getal} (1999) have carried out very large simulations
(containing $4\times 10^7$ particles) and suggest that the PS
formalism may under-predict the density of the rarest peaks by a
factor of about 10. Their observed halo abundance can be well fit by a
modified PS formula where the critical overdensity $\delta_c$ varies
with redshift and fluctuation amplitude via
\begin{equation}
\delta_c=1.685\left[\frac{0.7(1+z)}{\sigma_8}\right].
\end{equation}
In the central panel of Figure 11
we show the result for
the likelihood function derived using this modified form for
$\delta_c$ (long-dashed line). The resulting likelihood function
differs little from that obtained using the standard PS formalism, and
in particular the case of Gaussian fluctuations with $\Omega_{\rm
m}=1$ is still ruled out at the $2\sigma$ level. From this discussion
we infer that uncertainties about the validity of the PS formalism in
the rare peak limit should not have a large effect on the
conclusions of this paper.

Finally, we show the effect of assuming a flat universe with a
cosmological constant, rather than the open case which we have
considered up to now. Figure 12
shows the likelihood
function marginalized over $\sigma_8$ for the cases $G=1$,
$\Omega_m=0.3$, and $\Omega_m=0.6$ for open and flat universes. The lambda
case typically predicts slightly more evolution than the open case if the
other parameters are held fixed. Consequently, the best fit values of
$\Omega_m$ are slightly lower, or in the non-Gaussian case, the best
fit value of $G$ is slightly higher.

\subsection{Combined results}

Having considered the importance of various sources of systematic
error, we now show the likelihood function derived by combining all of
the cluster data at different redshifts. In Figure 13
we show
the likelihood function (marginalized over $\sigma_8$) for the cases
$G=1$, $\Omega_m=1.0$ and $\Omega_m=0.3$, using the combination of the
Markevitch, Henry, $0.5<z<0.65$ and $0.65<z<0.9$ datasets. We show the
results for open and flat universes together with two additional
analyses for the open case. The first, denoted ``Low $\Omega$'',
incorporates the major systematic uncertainties favoring a low value
of $\Omega_m$. The second, denoted ``High $\Omega$'', incorporates
the major systematic uncertainties favoring a high value of
$\Omega_m$. For the ``Low $\Omega$'' model, we replace the
Markevitch data at low redshift with the Henry and Arnaud sample,
without a volume limit, and we adopt a lower temperature bound of
$T_{\rm min}=12.0$ keV for our analysis of the $0.65<z<0.9$ sample. For
the ``High $\Omega$'' model, we multiply the Press-Schechter
prediction for the low redshift sample by a factor $F_1=0.75$, and the
predictions for the $0.5<z<0.65$ and $0.65<z<0.9$ samples by a factor
$F_2=1.25$. We also adopt a normalization for the mass-temperature
relationship of $\beta=1.3$, use a temperature for MS1054.5-0321 of
$10.1$keV, and adopt a lower temperature bound of $T_{\rm min}=6.3$
for the analysis of the $0.65<z<0.9$ sample. The results from the
``High $\Omega$'' and ``Low $\Omega$'' analyses span a reasonable
range of systematic uncertainty in the true likelihood function.

\vbox{%
\begin{center}
\leavevmode
\hbox{%
\epsfxsize=7.5cm
\epsffile{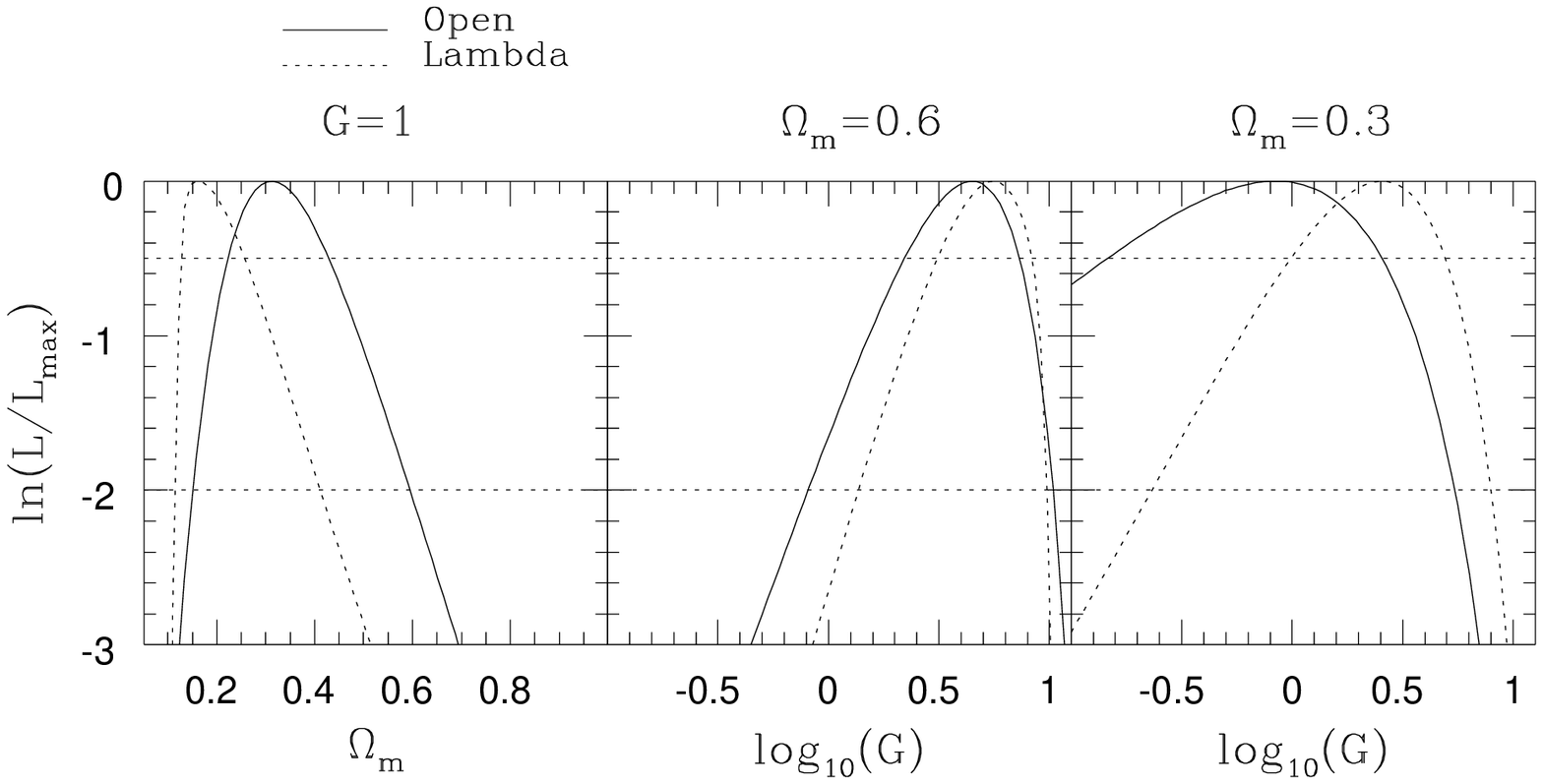}}
\begin{small}
\figcaption{\small 
Likelihood (marginalized over $\sigma_8$) for the Markevitch data and
the $0.65<z<0.9$ data, for open and flat universes.}
\end{small}
\label{fig-all_lambda}
\end{center}}

For interpreting most of the results in this paper, we have made the
simplifying assumption that the likelihood levels corresponding to a
given confidence limit are just the levels that would be appropriate
for a Gaussian distribution. In the Gaussian case the $1\sigma$ limit
is $\ln(L/L_{\rm max})>-0.5$ and the $2\sigma$ limit is $\ln(L/L_{\rm
max})>-2$. Clearly some of our likelihood functions are not well
approximated as a Gaussian distribution, and for this final analysis
we quote precise limits for each likelihood curve, assuming uniform
priors in $\Omega_{\rm m}$ and $\log_{10} (G)$, with the parameters
restricted to lie in the range plotted. We find that $1\sigma$
thresholds are within the range $-0.57<\ln(L/L_{\rm max})<-0.48$ and
the $2\sigma$ thresholds are within the range $-2.07<\ln(L/L_{\rm
max})<-1.93$ (that is, very close to the Gaussian levels) for almost
all the curves in Figure 13.
The only exceptions are the Open
and ``Low Omega'' models in the $\Omega_{\rm m}=0.3$ case, for which
the $1\sigma$ limit is $L/L_{\rm max}>-0.23$ and the $2\sigma$ limit
is $L/L_{\rm max}>-1.1$, and the ``High Omega'' model in the
$\Omega_{\rm m}=0.3$ case, for which the $1\sigma$ limit is $L/L_{\rm
max}>-0.23$, and the $2\sigma$ limit is $L/L_{\rm
max}>-1.75$.

To summarize the results of the full likelihood analysis, we find that
for a Gaussian universe, the best fit value of $\Omega_m$ is in the
range $\Omega_m\simeq 0.4\pm{0.25}$, with $\Omega_m=1$ inconsistent at
the $2\sigma$ level independent of almost all sources of systematic
error. The Gaussian, $\Omega_m=1$ case can only be reconciled at the
$2\sigma$ level by a conspiracy of several systematic effects working
in the right direction. If we assume that $\Omega_m=1$, then the
predicted degree of cluster evolution can be reconciled with the
observations if $G>2.0$, with the best fit value being $G\simeq6.5\pm
2.0$. Under the assumption that $\Omega_m=0.3$, Gaussianity is always
consistent with the data, but a wide range of non-Gaussian models also
fit, with all values $G<4$ (or $G<6$ in the lambda case) acceptable at
the $2\sigma$ level. These conclusions are robust to a wide range of
systematic uncertainties, and independent of whether the universe is
open or flat.

\vbox{%
\begin{center}
\leavevmode
\hbox{%
\epsfxsize=7.5cm
\epsffile{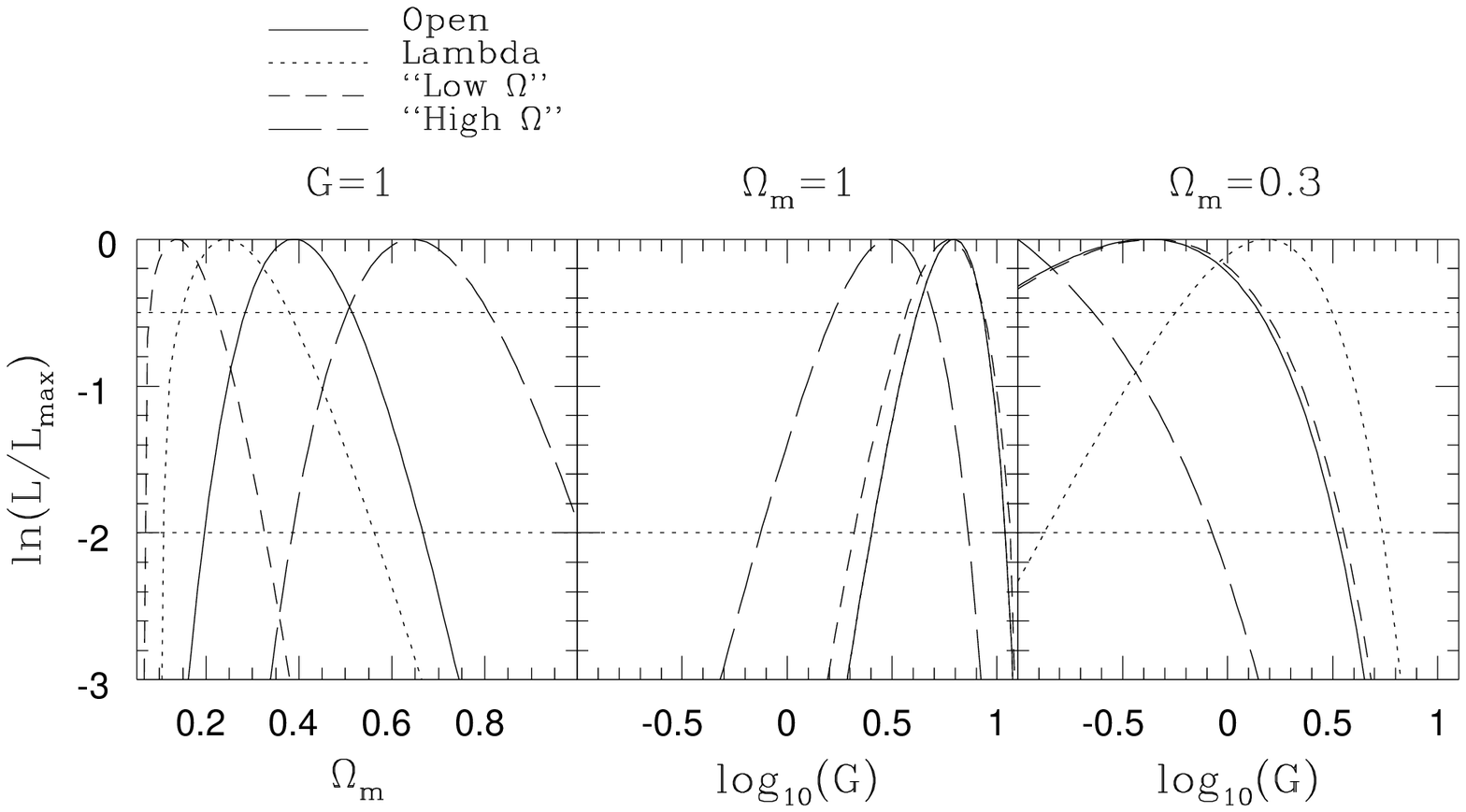}}
\begin{small}
\figcaption{\small  
Likelihood (marginalized over $\sigma_8$) for all datasets combined,
for an open universe, a lambda universe, and for ``Low $\Omega$'' and
``High $\Omega$'' analyses which incorporate a number of systematic
effects favoring low and high values of $\Omega_m$, respectively.}
\end{small}
\label{fig-all}
\end{center}}

We discuss briefly the degree of non-Gaussianity which these $G$
values represent. Figure 14
shows probability distribution
functions with $G=1$, $G=2.5$ and $G=9$. We see that the Gaussian and
the $G=2.5$ cases are barely distinguishable, with the differences
only becoming apparent for the rare tail of fluctuations. We reiterate
that the level of non-Gaussianity represented by the $G=2.5$ PDF is
sufficient to reconcile an $\Omega_m=1$ universe with the observed
degree of cluster evolution. The $G=9$ PDF differs from the Gaussian
case more obviously over its entire range, not just for the rare tail
of perturbations. We can also relate the parameter $G$ to some more
familiar characterizations of non-Gaussianity. In Figure 15,
we show the relationship between skewness $S$, kurtosis $K$, and
non-Gaussianity parameter $G$ for the log normal family of PDFs we are
considering. Since the PDF is normalized to have mean zero and
standard deviation one, the skewness is just
\begin{equation}
S=\int_{-\infty}^{\infty} y^3 P(y) \,dy. 
\end{equation}
while the kurtosis is
\begin{equation}
K=\int_{-\infty}^{\infty} y^4 P(y) \,dy \, -3. 
\end{equation}
The $G=2.5$ case has a skewness of $0.3$, and a kurtosis of $0.1$,
while the $G=9.0$ case has a skewness of $1.2$ and a kurtosis of
$2.6$. Some typical $G$ values for physical models are: cosmic strings
-- $G\simeq 5$ (RB00), cosmic textures
-- $G\simeq 14$ (Park, Spergel \& Turok\markcite{PST} 1991), Peebles
ICDM $G\simeq 15$ (RB00\markcite{RB}). The latter
two models would therefore appear to be ruled out if the matter
density of the universe is of order $0.3$.

\vbox{%
\begin{center}
\leavevmode
\hbox{%
\epsfxsize=7.5cm
\epsffile{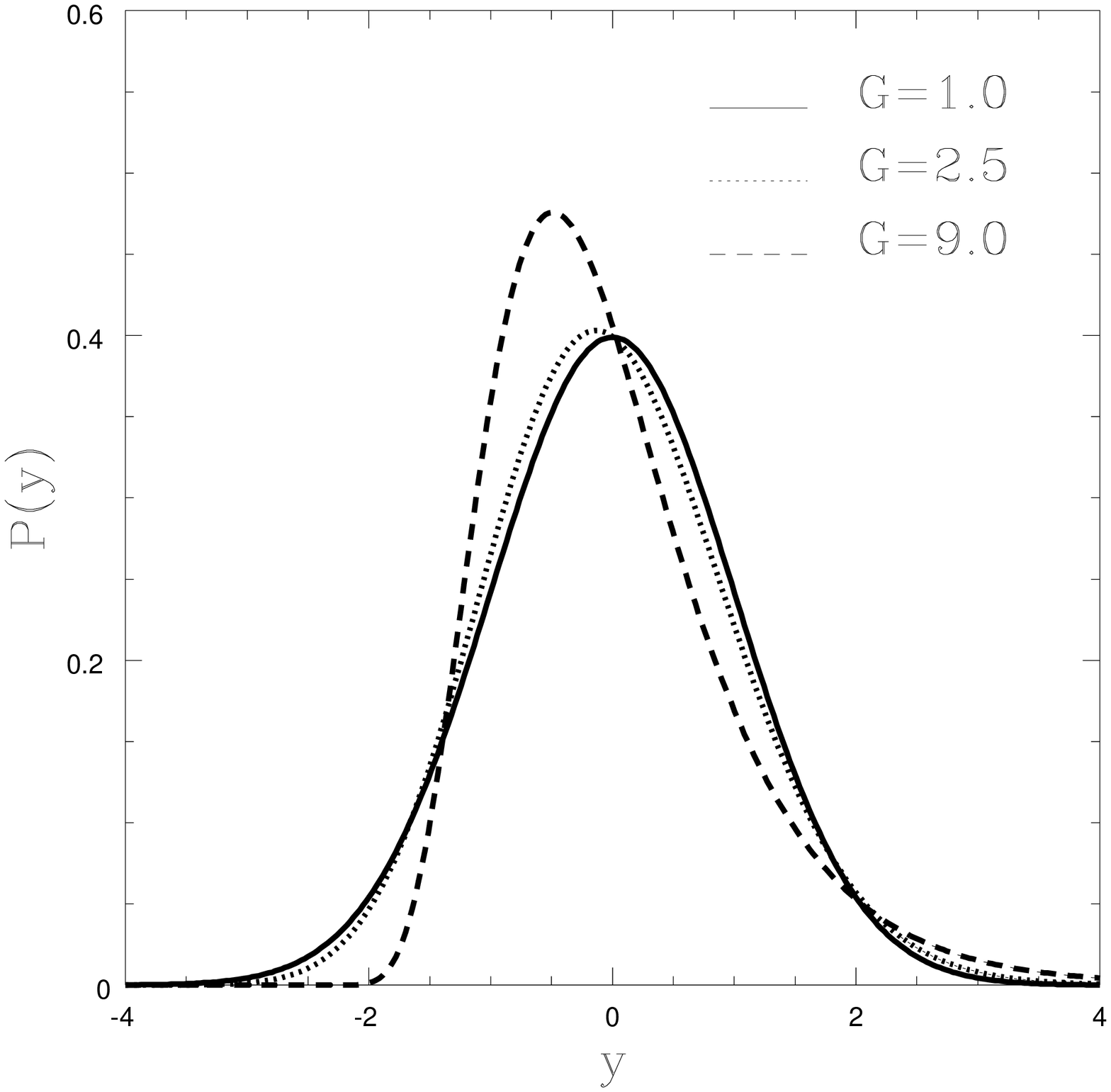}}
\begin{small}
\figcaption{\small 
One Gaussian and two non-Gaussian PDFs, with non-Gaussianity parameters
as labeled.} 
\end{small}
\label{fig-pdf}
\end{center}}

From this discussion, we see that the level of non-Gaussianity
required to reconcile an $\Omega_m=1$ universe with existing cluster
evolution data is relatively small, and less than the amount predicted
by many well motivated non-Gaussian models. One promising scenario
which could `save' $\Omega_m=1$ is a special form of ``hybrid
inflation'', where an initial spectrum of adiabatic Gaussian
fluctuations is further perturbed by the evolution of a network of
cosmic defects (Contaldi, Hindmarsh \& Magueijo 1999; Battye \&
Weller\markcite{BW} 1998). The level of non-Gaussianity in such a
model would be lower than that resulting from the action of the defect
network alone. Of course, for this model to work it would also have to
overcome a number of independent arguments against a critical density
universe (see e.g. Bahcall \& Fan\markcite{BFb} 1998b).

\vbox{%
\begin{center}
\leavevmode
\hbox{%
\epsfxsize=7.5cm
\epsffile{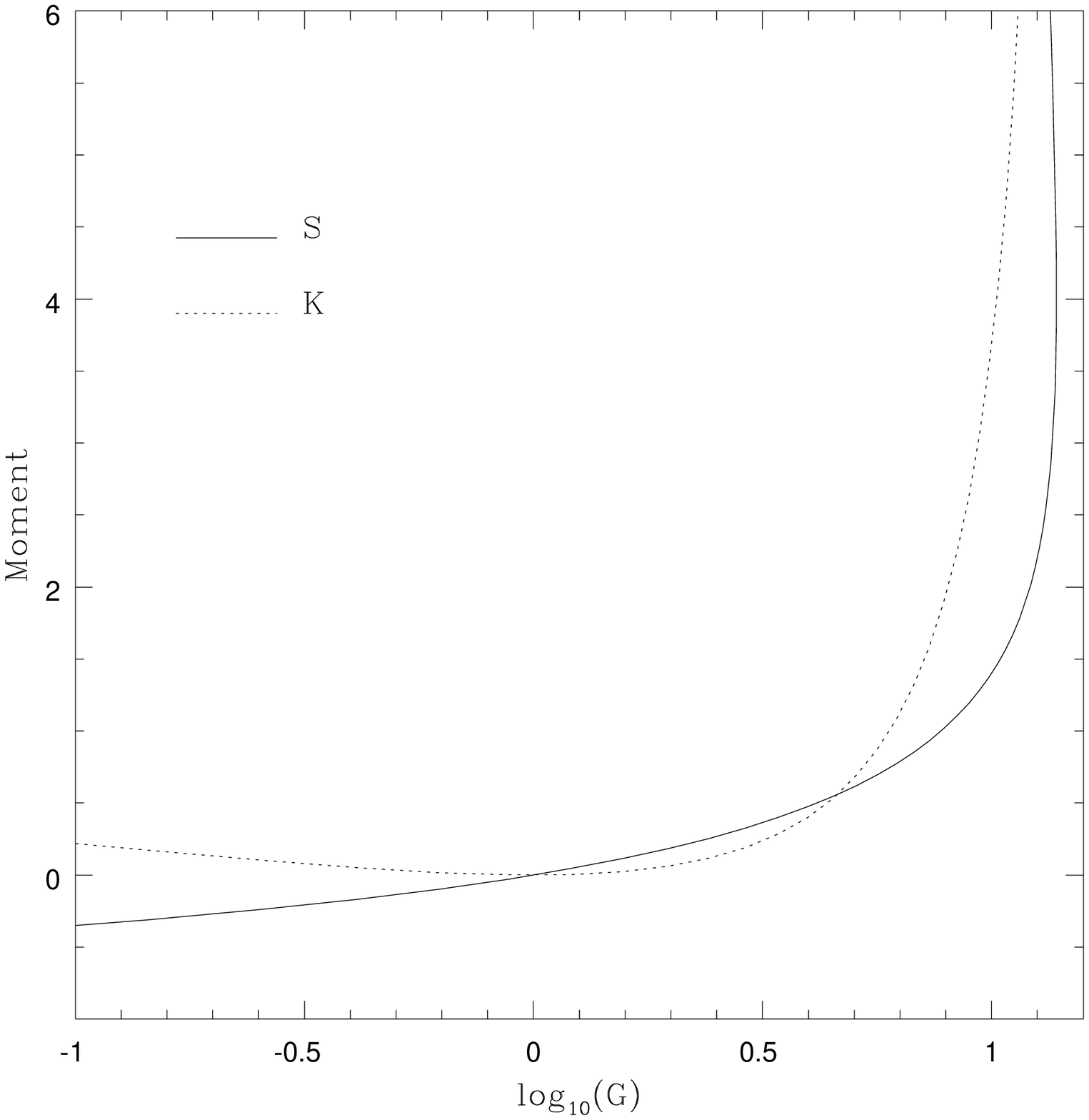}}
\begin{small}
\figcaption{\small 
Relationship between skewness and non-Gaussianity parameter $G$
for the log-normal family of PDFs.} 
\end{small}
\label{fig-skew}
\end{center}}

As one last point, we comment on our particular choice of the
parameter $G$ as a quantifier of non-Gaussianity. Let us consider a
more general non-Gaussianity statistic, $G_n$, defined as
\begin{equation}
\label{eqn-Gn}
G_n=2\pi \frac{\int_n^\infty P(y) dy}{\int_n^\infty e^{-y^2/2} dy}.
\end{equation}
$G_n$ is the probability of obtaining a peak of height $n\sigma$ or
higher, relative to that for a Gaussian distribution. For the bulk of
the discussion in this paper we make use of the parameter
$G=G_3$. Figure 16
shows the parameters $G_1$, $G_2$,
$G_3$, $G_4$ and $G_5$ as a function of $G$, for the case of a
log-normal PDF. Although the plot has been produced using a log-normal
distribution, an almost identical plot would result for a $\chi^2$
PDF. We have found that log-normal and $\chi^2$ distributions with the
same value of $G_3$ give very similar predictions for the cluster
abundance, at least over the range of masses and redshifts discussed
here. Since log-normal and $\chi^2$ PDFs with the same value of $G$
have very similar values of $G_2$, $G_4$ and $G_5$, any of these
parameters would also be an acceptable choice for parameterizing
non-Gaussianity. We concentrate on $G_3$ since the collapse threshold
of a typical cluster in the local universe is about $3\sigma$, so that
the value of $G_3$ has a simple physical interpretation. By contrast,
the parameter $G_1$ would not be a useful quantifier of
non-Gaussianity, since its value is almost independent of $G$. In
extreme cases where the PDF is not accurately fit by a log-normal
or $\chi^2$ distribution, the use of the single parameter $G$ may be
inadequate, and it may be necessary to consider the full form of the
distribution.

\vbox{%
\begin{center}
\leavevmode
\hbox{%
\epsfxsize=7.5cm
\epsffile{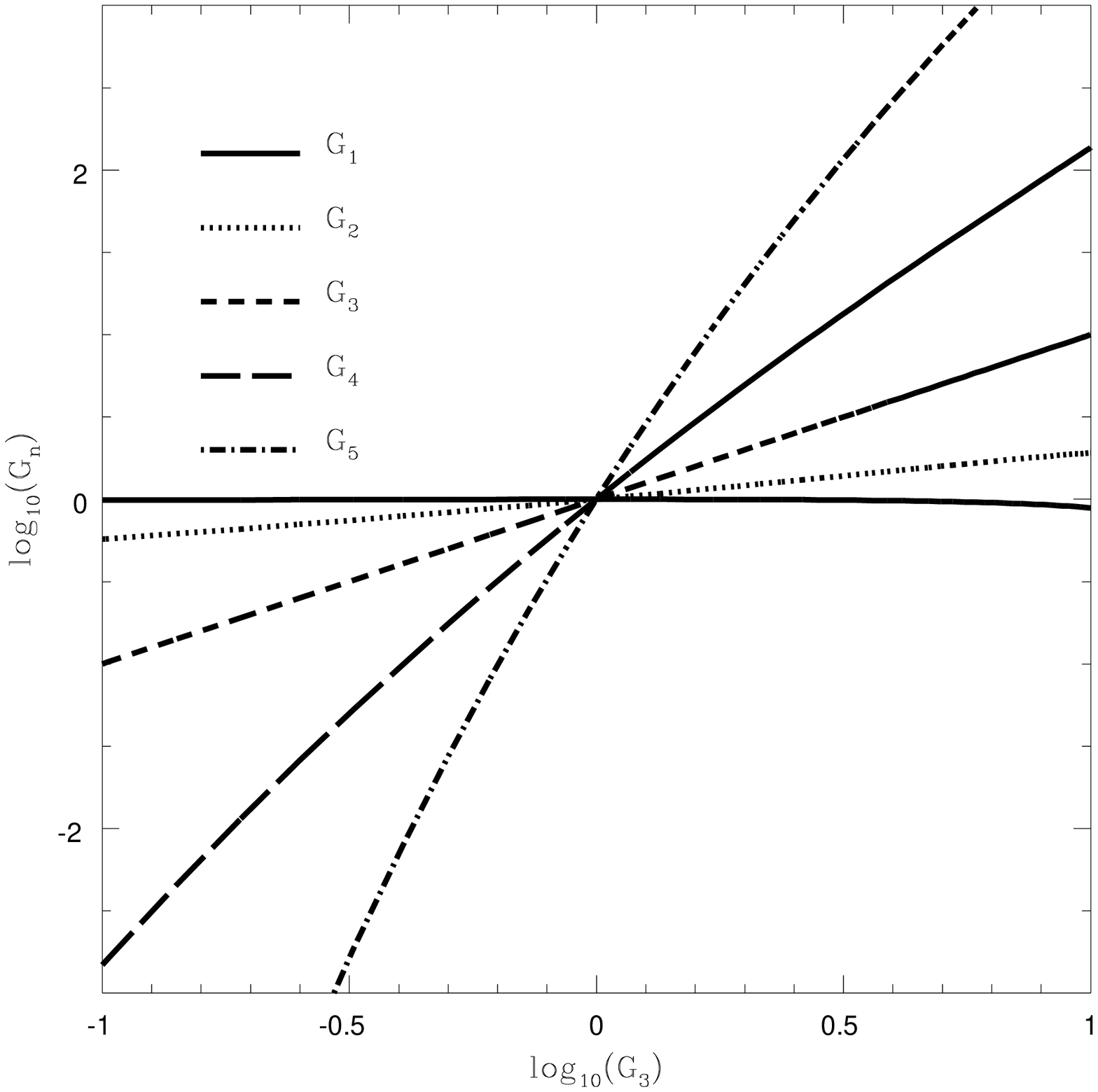}}
\begin{small}
\figcaption{\small 
Relationship between various non-Gaussianity parameters $G_n$ (see
equation~\ref{eqn-Gn}) and our preferred non-Gaussianity parameter $G$,
for a log-normal PDF. Results for a $\chi^2$ PDF are almost identical.} 
\end{small}
\label{fig-tails}
\end{center}}

\subsection{Future data}
\label{sec-future}
We have seen that existing cluster evolution data strongly disfavors
an $\Omega_m=1$ universe with Gaussian fluctuations, preferring either
a low value of $\Omega_m$, or a high value of $G$. An independent
measure of $\Omega_m$ (from CMB data and supernovae observations for
example) would therefore allow us to draw conclusions about the degree
of non-Gaussianity in the universe. Let us consider three different
models which are roughly degenerate with respect to current cluster
data: a model with $\Omega_m=1.0$, $G=8.0$, $\sigma_8=0.35$, a model
with $\Omega_m=0.6$, $G=4.5$, $\sigma_8=0.525$, and a model with
$\Omega_m=0.3$, $G=1.0$, $\sigma_8=0.85$. The likelihood function of
these models relative to the observational data considered above is
shown in Figure 17.
To evaluate these likelihood
functions, we have assumed that $\Omega_m$ is measured to within
$0.05$ for the $\Omega_m<1$ cases, and then marginalized over this
uncertainty. For the $\Omega_m=1.0$ model, assuming that we have an
unambiguous determination of the true value of $\Omega_m$, the current
cluster data gives a clear detection of the non-Gaussianity of the
fluctuations, as discussed above. For the $\Omega_m=0.6$ model, the
current observational data is unable to detect the non-Gaussianity,
with Gaussian fluctuations allowed at the $2\sigma$ level. For the
$\Omega_m=0.3$ case, Gaussian fluctuations are consistent with the
data, with the non-Gaussianity parameter constrained to be $G<4.0$ at
the $2\sigma$ level.

\vbox{%
\begin{center}
\leavevmode
\hbox{%
\epsfxsize=7.5cm
\epsffile{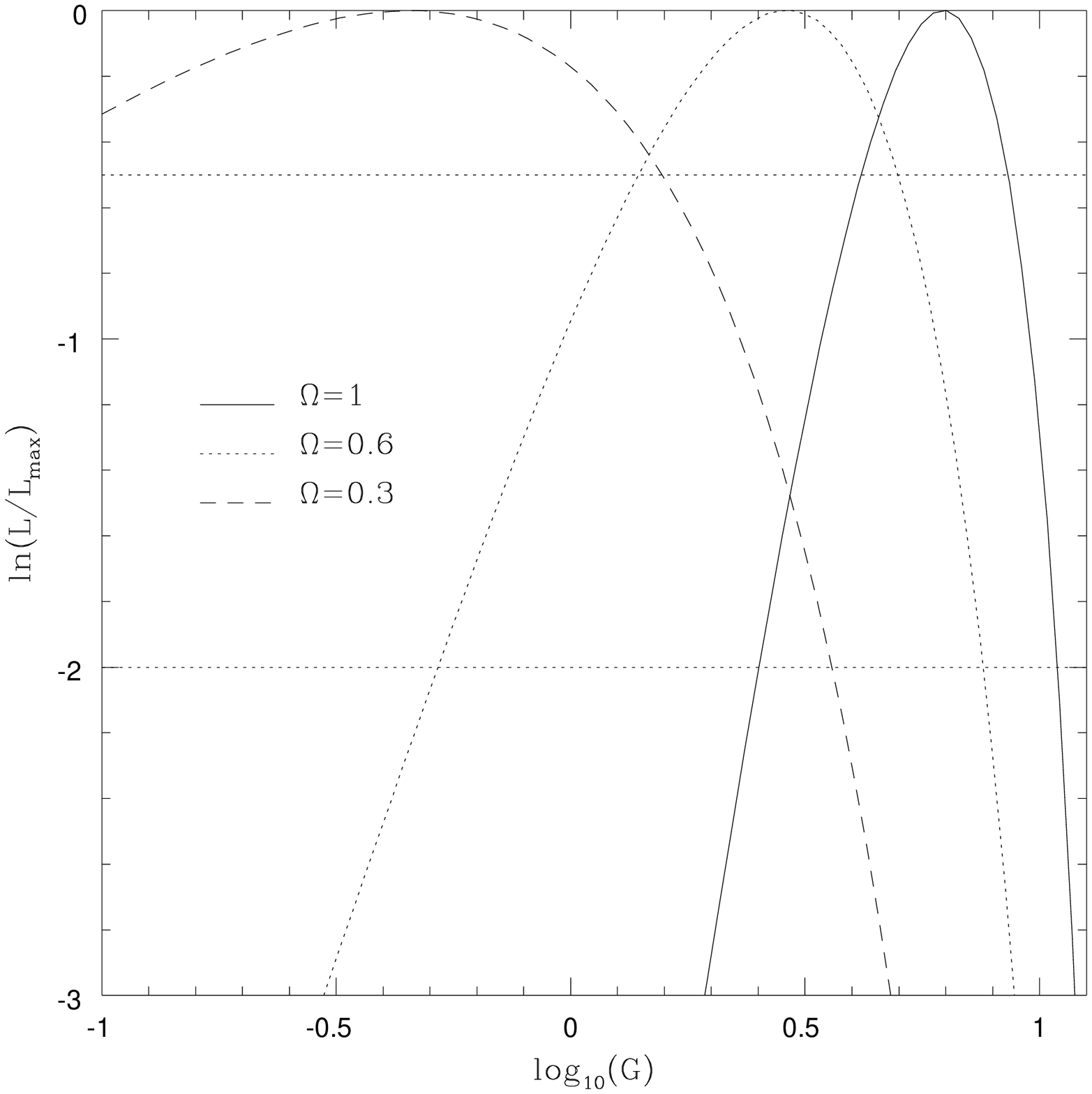}}
\begin{small}
\figcaption{\small 
Likelihood (marginalized over $\sigma_8$) for the three models
discussed in section~\ref{sec-future} and the combination of the
Markevitch and $0.65<z<0.9$ datasets. The models are labeled by the value of
$\Omega_m$, and analyzed assuming that the true value of $\Omega_m$ has
been measured precisely in each case.} 
\end{small}
\label{fig-omega0.6}
\end{center}}

For each of the models just mentioned, we have also computed likely
confidence limits which could be achieved by a considerable but
realistic increase in the volume of cluster evolution data. For each
model, we have generated mock realizations of the Markevitch
catalogue, and of a complete EMSS catalogue, assuming that
temperatures are measured for all clusters within the EMSS survey for
which $T>6.3$keV and $z<0.9$. To generate these catalogues, we split
the surveys into a number of redshift slices (one slice for the Markevitch
data, with the same redshift and flux limits and the same sky
coverage, and four slices for the EMSS data with $0.3<z<0.4$,
$0.4<z<0.5$, $0.5<z<0.65$, and $0.65<z<0.9$, assuming the sky coverage
and flux limits of the EMSS survey). For each slice, we compute the
mean volume surveyed for each cluster temperature, assuming the
fiducial LT relationships given above. We then generate a Poisson
sample of clusters satisfying the $n(T)$ law for the model in question,
computed at the median redshift of the sample.

We now assume that the true value of $\Omega_m$ has been measured by
some independent method (to plus or minus 0.05, except for the
$\Omega_m=1$ case, where we assume an exact determination) and ask
how well the mock catalogues are able to constrain the non-Gaussianity
parameter $G$. Results for the likelihood function (marginalized over
$\sigma_8$, which is unknown, and the uncertainty in $\Omega_m$) are
shown in Figure 18.
In the $\Omega_m=1$ case
the improved data considerably reduces the errors on the determination
of $G$, and in the $\Omega_m=0.6$ case an unambiguous detection of
non-Gaussianity would be possible at the $2\sigma$ level. In the
$\Omega_m=0.3$ case, the allowed range of $G$ is not significantly
reduced, with the limit still being $G<4$ at the $2\sigma$ level.

\vbox{%
\begin{center}
\leavevmode
\hbox{%
\epsfxsize=7.5cm
\epsffile{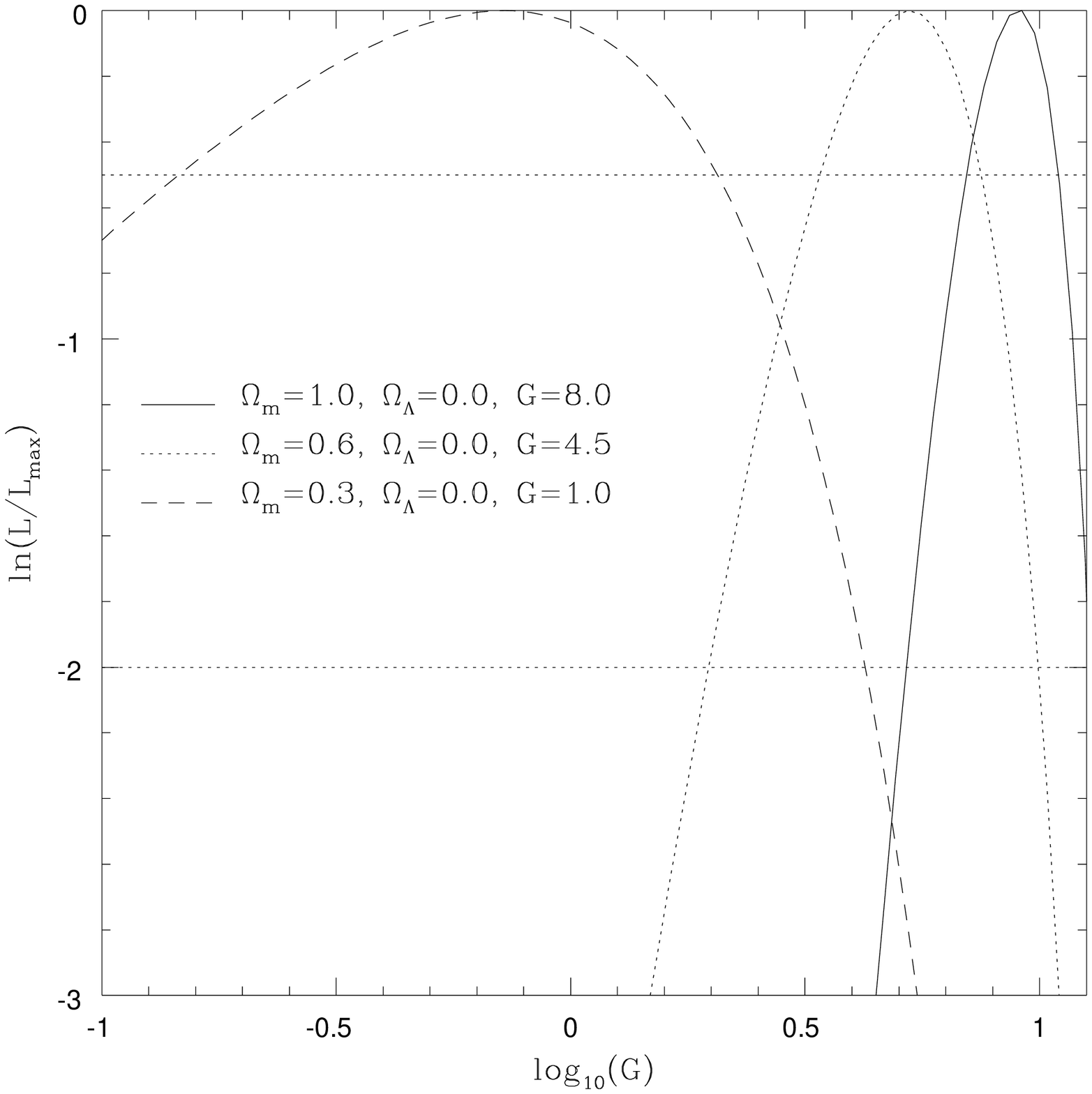}}
\begin{small}
\figcaption{\small 
Mock likelihood functions (marginalized over $\sigma_8$) for the three models
discussed in section~\ref{sec-future} and the combination of the
Markevitch and EMSS {\it mock} datasets. The models are labeled by the value of
$\Omega_m$, and analyzed assuming that the true value of $\Omega_m$ has
been measured (plus or minus errors of 0.05 in the $\Omega_m<1$
cases).}
\end{small}
\label{fig-omega_marginalize}
\end{center}}

From the above discussion, we see that even the considerable
improvement in the quantity of cluster data discussed above would not
significantly improve constraints on non-Gaussianity if $\Omega_m$ is
really of order 0.3. The reason cluster evolution does not give us
such strong constraints in this case is that in a low density
universe, clusters are not such rare events. For this reason, they do
not probe the high-$\sigma$ tail of the PDF, which is where Gaussian
and non-Gaussian PDFs tend to differ most significantly. This fact is
illustrated in Figure 19,
where we show the cumulative
number density of clusters expected at redshifts $z=0.05$, $z=0.8$,
$z=1.5$ and $z=2.0$ for two models of structure formation (Gaussian --
$G=1.0$, $\sigma_8=0.85$, and non-Gaussian -- $G=3.0$,
$\sigma_8=0.75$), both with $\Omega_m=0.3$, which are nearly
degenerate with respect to the real and mock data discussed above. We
see that the degree of evolution between $z=0.05$ and $z=0.8$ begins
to differ considerably between the two models only for cluster
temperatures larger than about 20 keV. However, clusters this hot are
so rare ($N_{>T}<10^{-10}h^3$Mpc$^{-3}$) that the surveys considered
above do not probe enough volume to place any constraints on the
number density. As we move to higher redshifts, lower temperature
clusters become rarer, many-$\sigma$ events, and the degree of
evolution to $z=1.5$ differs considerably between the two models for
cluster temperatures of order 15 keV. For the non-Gaussian model
discussed above, the number density of such clusters at redshift
$z=1.5$ is $N_{>15{\rm keV}}\simeq10^{-9}h^3$Mpc$^{-3}$, meaning that
a survey sensitive to clusters in the range $1.25<z<1.75$ would have
to cover approximately 1000 square degrees to find an average of one
$T>15$ keV cluster. The existence of such a cluster would however be
ten times less likely in a Gaussian universe, where $N_{>15{\rm
keV}}\simeq10^{-10}h^3$Mpc$^{-3}$. We conclude, therefore, that if the
matter density of the universe does indeed turn out to be of order
$z=0.3$, much deeper cluster data will be required to improve
constraints on non-Gaussianity, with surveys covering 1000 or more
square degrees capable of detecting hot clusters ($z>15$ keV) at
redshifts $z\ge 1.5$. A catalogue based on serendipitous cluster
detections from the forthcoming XMM satellite could hope to cover such
an area to sufficient depth (Romer\markcite{R98} 1998).

\vbox{%
\begin{center}
\leavevmode
\hbox{%
\epsfxsize=7.5cm
\epsffile{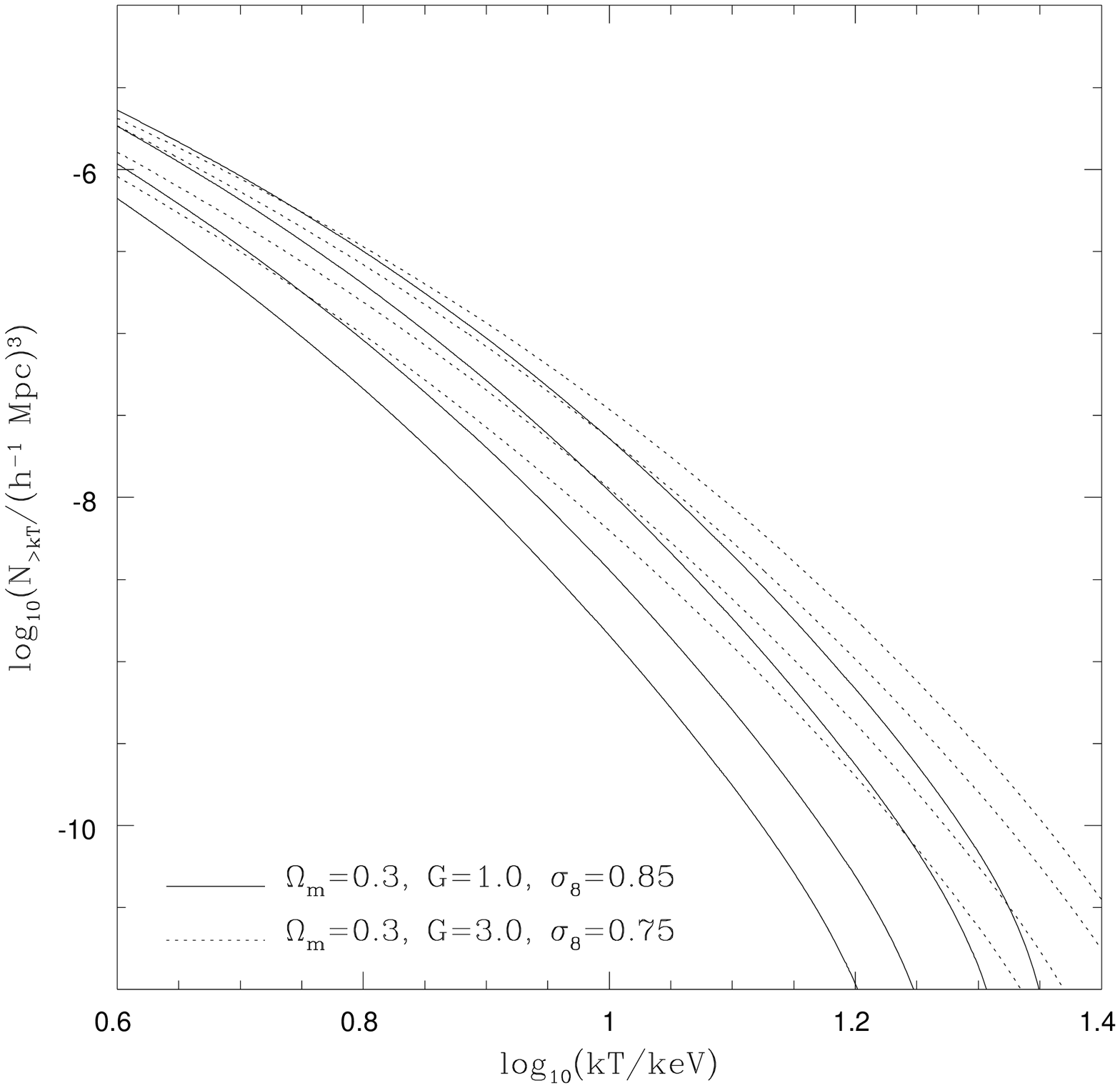}}
\begin{small}
\figcaption{\small Predicted cluster abundances at $z=0.05$, $z=0.8$, $z=1.5$,
$z=2$ (highest to lowest curves, respectively), for a Gaussian and a
non-Gaussian model, as labeled, with $\Omega_m$=0.3. 
}
\end{small}
\label{fig-curves}
\end{center}}

We make one final point on the degree to which future data will be
able to constrain non-Gaussianity. From the above discussion, it is
clear that the ability to detect non-Gaussianity depends on the
detection of clusters arising from rare peaks in the primordial
distribution. In addition to carrying out deeper observations, it will
obviously be important to extend theoretical tests of the non-Gaussian
Press-Schechter formalism to larger volumes, and to investigate in
detail any deviations from the PS formulae which exist in the rare
peak limit.

\section{Conclusions}
\label{sec-conclusions}
We have made use of a modified version of the Press-Schechter
formalism to examine the constraints on non-Gaussianity which can be
derived from observations of evolution of the galaxy cluster number
abundance. Our results are summarized in Figure 13, 
which shows likelihood functions for the combined cluster datasets for the
cases $G=1$ (Gaussian fluctuations), $\Omega_m=1$, and
$\Omega_m=0.3$. For an open universe with Gaussian fluctuations, the
maximum likelihood value of the matter density on the basis of current
observations is $\Omega_m = 0.4\pm{0.25}$, with the
case $\Omega_m=1$ inconsistent with the data at the $2\sigma$
level. If we assume an $\Omega_m=1$ universe with non-Gaussian
fluctuations, the best fit value of the non-Gaussianity parameter is
in the range $G=6.5\pm 2$, with $G>2.0$ required for consistency with
the data at the $2\sigma$ level. The degree of non-Gaussianity
required to `save' the critical density universe is therefore
relatively small, and could easily be realized by physically motivated
models such as the combination adiabatic and defect seeded
perturbations predicted by some Hybrid inflation models. If we assume
an $\Omega_m=0.3$ universe, the non-Gaussianity parameter is
constrained to be $G<4$ ($G<6$ in the lambda case) at the $2\sigma$
level, implying that some well motivated non-Gaussian models, such as
textures ($G\simeq 14$) and Peebles ICDM ($G\simeq 15$) would be
inconsistent with the data. These conclusions are robust to a wide
range of sources of systematic uncertainty. An $\Omega_m=1$ universe
with Gaussian fluctuations could only be reconciled with the data if a
conspiracy of several systematic errors were all to modify our
conclusions in the right direction. In particular, our results are
unaffected by possible systematic errors in the Press-Schechter
prediction for the cluster abundance in non-Gaussian models, which
RB00\markcite{RB} has shown to be less than 25$\%$. Our conclusions
are also largely independent of whether the universe is open or flat
(slightly lower values of $\Omega_m$ or higher values of $G$ are
preferred in the cosmological constant case). Since the non-Gaussian
Press-Schechter formalism requires only the probability distribution
function of primordial fluctuations as input, our analysis is
independent of any uncertain features of the non-Gaussian physics in
specific models. One source of uncertainty which does deserve further
attention is the possibility that the PS formalism could break down in
the very rare halo limit, as suggested by recent very large
simulations (for example Governato et al. 1999). Our preliminary
analysis in section~\ref{sec-results} suggests that this effect should
not be too important for the range of cluster masses and redshifts
discussed here, but this conclusion should be tested using larger
simulations in the non-Gaussian case.

The techniques discussed here allow us to constrain primordial
non-Gaussianity in the universe, provided we have an independent
measurement of the matter density $\Omega_m$. We can realistically
expect to gain such a measurement with high precision from the
combination of upcoming CMB and supernovae data. If the matter density
is measured to be $\Omega_m=1$ then the current cluster evolution data
represents a detection of non-Gaussianity. If the matter density is
measured to be $\Omega_m=0.6$ or lower then both Gaussian and
non-Gaussian fluctuations are consistent with current cluster
evolution data.

Our results for the Gaussian case agree well with those of Bahcall \&
Fan\markcite{BF} (1998a) who find $\Omega_m=0.2^{+0.3}_{-0.1}$ and Eke
et al.\markcite{ECFH} (1998) who find $\Omega_m=0.45\pm0.2$, each
using similar data to that considered here. Our results disagree
however with those of Blanchard \& Bartlett (1998), who find evidence
in favor of an $\Omega_m=1$ universe by combining the EMSS and ROSAT
samples. As for non-Gaussian models, our results are consistent with
those of Willick\markcite{Willick} (1998) in that both studies allow
for a significant degree of non-Gaussianity if
$\Omega_m\simeq0.3$. Our analysis however does not significantly
disfavor Gaussianity in this case. This difference is probably due to
the fact that we include more clusters and allow for more sources of
systematic error, particularly in the low-redshift normalization. Our
results are consistent with those of RGS98 and Koyama et al. (1999)
provided that the matter density of the universe is low.  In
particular, RGS98 found that a significant amount of non-Gaussianity
($G\simeq4.0^{+3.6}_{-2.0}$) was required to reconcile cluster
abundance and cluster correlation data in the $\Omega_m=0.3$ case. We
caution the reader however that the arguments used in these last two
papers depend on a model for halo correlations in non-Gaussian models
whose agreement with N-body simulations is currently uncertain. On the
other hand, the arguments used in the current work, and that of
Willick (1998), are based on a model for halo abundance in
non-Gaussian models which has been shown to agree well with that
observed in simulations.

We have also investigated the improvement to constraints on
non-Gaussianity which we could expect from a realistic increase in the
quantity of data. If the matter density of the universe is measured to
be high ($\Omega_m$ greater than of order 0.6), then a moderate
increase in the amount of cluster evolution data will allow a definite
detection of non-Gaussianity. However, if the matter density is
measured to be low, then a substantial increase in the quantity of
cluster evolution will be required in order to significantly improve
upon the limits derived here. The reason that constraints are weaker
in the $\Omega_m=0.3$ case is that clusters at moderate redshifts
($z\simeq 0.9$) are no longer particularly rare events, while the
strongest constraints on Gaussianity come from probing the rarest tail
of fluctuations. The best prospect for constraining non-Gaussianity in
this case is to carry out deep surveys capable of detecting hot
clusters ($T>15$ keV) at high redshift ($z\ge 1.5$). Such surveys would
be able to distinguish a non-Gaussian universe with $G=3$ from a
universe with Gaussian fluctuations if the area covered was of order
1000 square degrees. Conducting such surveys is a challenge which will
deserve considerable attention if the matter density is indeed
confirmed to be low.

\section{Acknowledgments}
We would like to thank Pat Henry for supplying updated information on
the Henry (1997) cluster sample. We would also like to thank Marc
Davis for helpful discussions.  This work has been supported in part
by NSF grant 9617168, and E.G. acknowledges partial support from
NASA AISRP (NAG-3941).

\newpage

\end{document}